\documentclass{aa}
\usepackage{natbib}
\usepackage[varg]{txfonts}
\usepackage{soul}
\usepackage{ltablex}
\usepackage{placeins}
\usepackage{subcaption}
\usepackage{lscape}
\usepackage{amsmath}
\usepackage{float}
\bibpunct{(}{)}{;}{a}{}{,} 
\usepackage{tikz}

\begin{document}
\title{J-PLUS: A first glimpse at spectrophotometry of asteroids - The MOOJa catalog}
\titlerunning{J-PLUS: A first glimpse at spectrophotometry of asteroids - The MOOJa catalog}

\author{David Morate\inst{1}
     \and Jorge Marcio Carvano\inst{1}
     \and Alvaro Alvarez-Candal\inst{2,3,1}
     \and M\'ario De Pr\'a\inst{4}
     \and Javier Licandro\inst{5,6}
     \and Andr\'es Galarza\inst{1}
     \and Max Mahlke\inst{7,8}
     \and Enrique Solano-M\'arquez\inst{8}
     \and Javier Cenarro\inst{9}
     \and David Crist\'obal-Hornillos\inst{9}
     \and Carlos Hern\'andez-Monteagudo\inst{9}
     \and Carlos L\'opez-Sanjuan\inst{9}
     \and Antonio Mar\'in-Franch\inst{9}
     \and Mariano Moles\inst{9}
     \and Jes\'us Varela\inst{9}
     \and H\'ector V\'azquez Rami\'o\inst{9}
     \and Jailson Alcaniz\inst{1}
     \and Renato Dupke\inst{1}
     \and Alessandro Ederoclite\inst{10}
     \and Claudia Mendes de Oliveira\inst{10}
     \and Laerte Sodr\'e Jr.\inst{10}
     \and Raul E. Angulo\inst{11}
     \and Francisco M. Jiménez-Esteban\inst{8}
     \and Beatriz B. Siffert\inst{12}
     \and J-PLUS collaboration}

  \institute{Observat\'orio Nacional, Coordenaç\~ao de Astronomia e Astrofísica, 20921-400 Rio de Janeiro, Brazil
  \and Instituto Universitario de F\'isica Aplicada a las Ciencias y las Tecnolog\'ias, Universidad de Alicante, San Vicent del Raspeig, E03080, Alicante, Spain
  \and
  Instituto de Astrof\'isica de Andaluc\'ia, CSIC, Apt 3004, E18080 Granada, Spain
  \and Florida Space Institute, University of Central Florida, Orlando, FL 32816, USA
  \and Instituto de Astrof\'isica de Canarias (IAC), C/V\'ia L\'actea s/n, 38205 La Laguna, Tenerife, Spain
  \and Departamento de Astrofísica, Universidad de La Laguna, 38205 La Laguna, Tenerife, Spain
  \and Université C\^ote d'Azur, Observatoire de la C\^ote d'Azur, CNRS, Laboratoire Lagrange, Bd. de l'Observatoire, CS 34229, 06304 Nice cedex 4, France
  \and Departmento de Astrofísica, Centro de Astrobiología (CSIC-INTA), ESAC Campus, Camino Bajo del Castillo s/n Villanueva de la Cañada, E-28692 Madrid, Spain; Spanish Virtual Observatory, Spain
  \and Centro de Estudios de Física del Cosmos de Aragón (CEFCA), Unidad Asociada al CSIC, Plaza San Juan 1, 44001 Teruel, Spain
  \and Instituto de Astronomia, Geofísica e Ciências Atmosféricas, Universidade de São Paulo, 05508-090, São Paulo, Brazil
  \and Donostia International Physics Centre (DIPC), Paseo Manuel de Lardizabal 4, 20018 Donostia-San Sebastian, Spain; IKERBASQUE, Basque Foundation for Science, 48013, Bilbao, Spain
  \and Campus Duque de Caxias, Universidade Federal do Rio de Janeiro, 25265-970, Duque de Caxias, RJ, Brazil}

\abstract 
{The Javalambre Photometric Local Universe Survey (J-PLUS) is an observational campaign that aims to obtain photometry in 12 ultraviolet-visible filters (0.3--1 $\mu$m) of $\sim$8\,500 deg$^2$ of the sky observable from Javalambre (Teruel, Spain). Due to its characteristics and strategy of observation, this survey will let us analyze a great number of Solar System small bodies, with improved spectrophotometric resolution with respect to previous large-area photometric surveys in optical wavelengths.} 
{The main goal of this work is to present here the first catalog of magnitudes and colors of minor bodies of the Solar System compiled using the first data release (DR1) of the J-PLUS observational campaign: the Moving Objects Observed from Javalambre (MOOJa) catalog.}
{Using the compiled photometric data we obtained very-low-resolution reflectance (photospectra) spectra of the asteroids. We first used a $\sigma$-clipping algorithm in order to remove outliers and clean the data. We then devised a method to select the optimal solar colors in the J-PLUS photometric system. These solar colors were computed using two different approaches: on one hand, we used different spectra of the Sun, convolved with the filter transmissions of the J-PLUS system, and on the other, we selected a group of solar-type stars in the J-PLUS DR1, according to their computed stellar parameters. Finally, we used the solar colors to obtain the reflectance spectra of the asteroids.}
{We present photometric data in the J-PLUS filters for a total of 3\,122 minor bodies (3\,666 before outlier removal), and we discuss the main issues of the data, as well as some guidelines to solve them.}
{}

\maketitle 


\section{Introduction}
\label{introduction}

The minor bodies that populate the Solar System are considered to be one of the most pristine materials within it, remnants of the processes that followed the condensation of the pre-solar nebula and the formation of the planets. Since then, these objects have undergone little or none geological and thermal transformations. However, they have experienced intense collisional events that affected their shape, size and surface composition, among other properties; they have also been exposed to different dynamical mechanisms, which transported these objects from their original locations to their current orbits. Thus, the study of minor bodies would provide answers to questions about the origin and evolution of the Solar System.


\begin{figure*}
\centering
\includegraphics[width=\hsize]{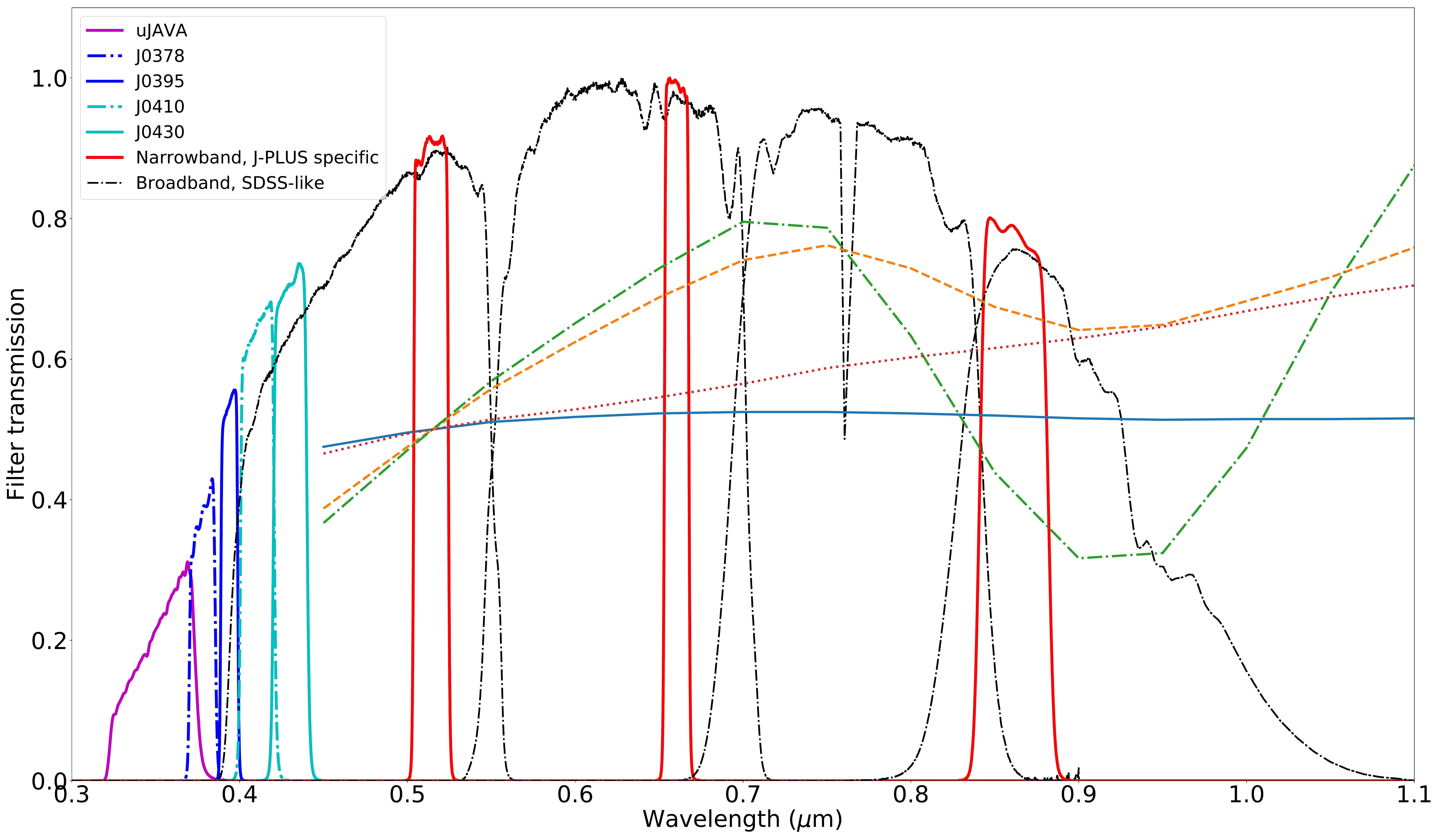}
\caption{J-PLUS filter system. The solid magenta line represents the $uJAVA$ filter; the blue adjacent curves (dotted and solid) represent the $J0378$ and $J0395$ filters, respectively; the following cyan curves (dotted and solid) represent the $J0410$ and $J0430$ filters; the three solid red lines are the narrow-band filters $J0515$, $J0660$, and $J0861$; these eight filters were designed specifically for J-PLUS. The remaining dashed black lines are the broad-band SDSS filters: from left to right, lines: $g$, $r$, $i$, $z$. For comparison, the spectral templates of four taxonomical classes (source: \protect\url{http://smass.mit.edu/busdemeoclass.html}) have been included: C (solid, blue), X (dotted, red), S (dashed, orange), and V (dash-dotted, green). The spectra have been shifted for better graphical representation, and normalized at 0.515 microns, which, out of the central wavelengths of every filter, is the nearest value to 0.55 microns, the most commonly used value for normalization of reflectance spectra of asteroids in the visible wavelength range. Note that the spectral templates do not cover the ultraviolet region (below 0.45 microns).}
\label{fig:filter_transmissions}
\end{figure*}

As pointed out by \cite{demeocarry2014}, the next step to improve our knowledge on this matter is to obtain spectra for a large sample of minor bodies. Spectroscopic observations of asteroids provide information on the surface composition of these objects. However, due to the observational methods, the number of spectra that are present in the literature is somehow limited: for visible wavelengths there are spectra available for $\sim$3000 objects (e.g. \citealt{busbinzel02,lazzaro13,deleon18}), while near-infrared spectra are available for only $\sim$1000 (see \citealt{busdemeo2009} and \citealt{mithneos19}).

Since the beginning of the 21st century, large-area surveys have become a powerful and usual tool to respond unanswered questions in different astronomic fields. Examples of these projects are SDSS \citep{york2000_sdss} and Pan-STARRS \citep{kaiser10_panstarrs} at visible wavelengths, and WISE \citep{wright2010_wise}, AKARI \citep{ishihara2010_akari}, 2MASS \citep{skrutskie2006_2mass}, VISTA \citep{sutherland15}, and UKIDSS \citep{lawrence07} at different infrared wavelengths. Given the vast fractions of the sky that these surveys cover in their observations, they pose the opportunity to analyze a large sample of Solar System objects (SSOs) which, although may not be the original goal, lie scattered within the obtained data.

Several works have taken advantage of the aforementioned surveys to compile large photometric datasets of minor bodies: \cite{ivezic2001} extracted data from the SDSS to create the SDSS Moving Object Catalog (MOC), whose last version contains data for more than 450\,000 objects\footnote{An updated version of the MOC catalogue can be found at \url{http://svo2.cab.inta-csic.es/vocats/svomoc}}; \cite{popescu16_movis}, using data from the VISTA Hemisphere Survey (VHS), created the Moving Objects from VISTA Survey (MOVIS) catalog with data for more than 50\,000 objects; \cite{mainzer11} obtained mid-infrared photometry for more than 150\,000 objects from the WISE survey, and produced the largest database (so far) of sizes and albedos of asteroids. These datasets have in common a limited number of filters (five in the SDSS and four in VISTA and WISE). However, even with only four or five filters, the minor bodies science production is extensive, due to the large number of observed objects. Some of the most relevant works derived from each of the previously mentioned surveys can be found in \cite{ivezic2002, parker2008, carvano2010, hasselman2011_dataset, masiero11, rivkin12, masiero13, alilagoa13, mainzer14, masiero14, hasselmann2015, licandro17, popescu18, morate18b}.


\begin{figure}[!h]
\centering
\includegraphics[width=\hsize]{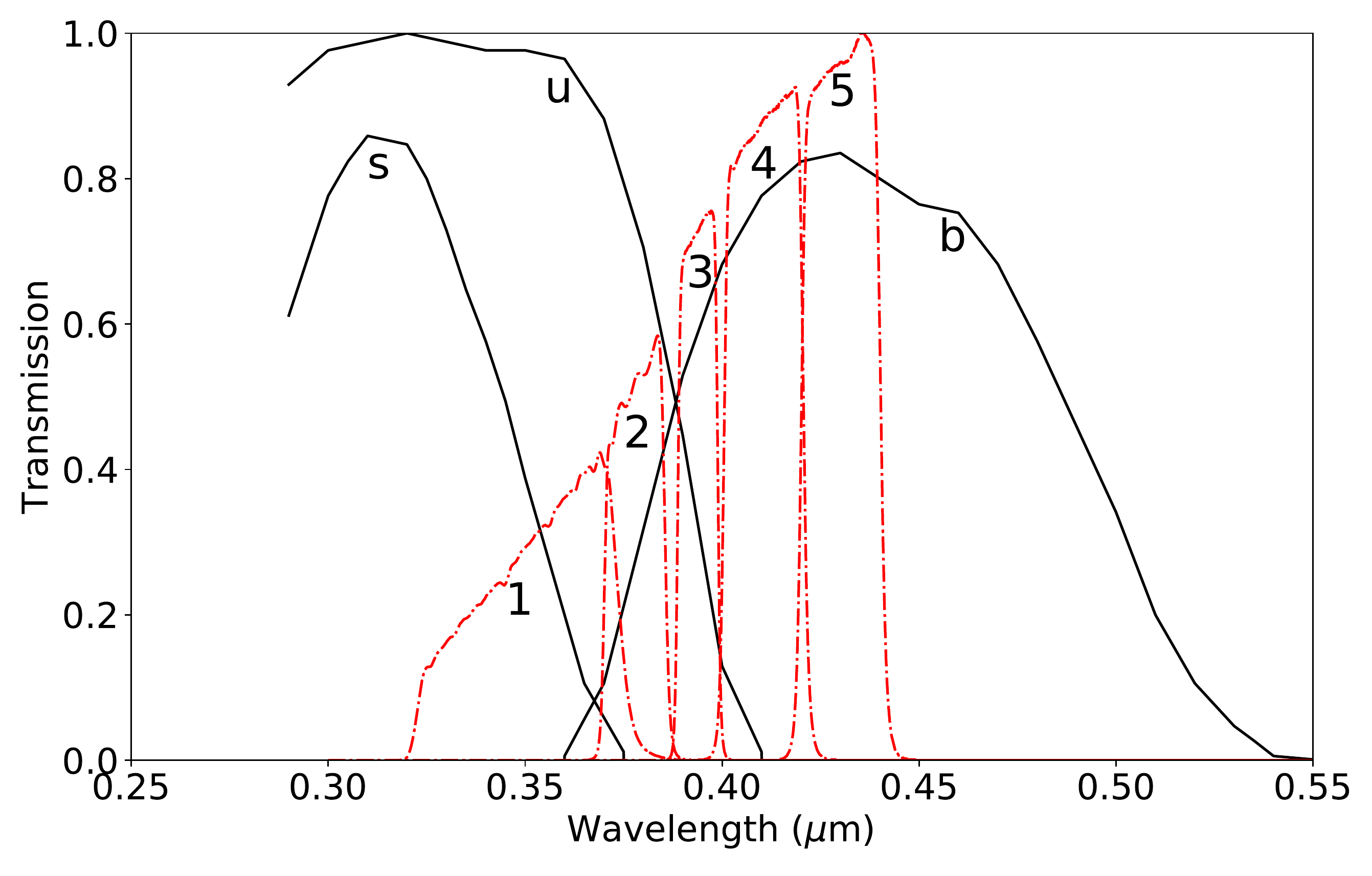}
\caption{Graphical comparison of the transmissions of the ECAS filters (black solid lines), and the J-PLUS filters (red dashdotted lines) in the ultraviolet region. The numbers correspond to: 1-$u$, 2-$J0378$, 3-J3095, 4-$J0410$, 5-$J0430$. ECAS transmissions are normalized to unity at the maximum of the u filter, and J-PLUS transmissions are normalized to unity at the maximum of the $J0430$ filter.}
\label{fig:jplus_vs_ecas}
\end{figure}


\subsection*{The Javalambre Photometric Local Universe Survey}

J-PLUS is a photometric sky survey which aims to cover $\sim$8\,500 deg$^2$, designed to observe and characterize galaxies and stars of the Milky Way halo, with a wide range of astrophysical applications \citep{cenarro19}. See also \url{https://www.j-plus.es/}. 


\begin{figure*}[h]
\centering
\includegraphics[width=\hsize]{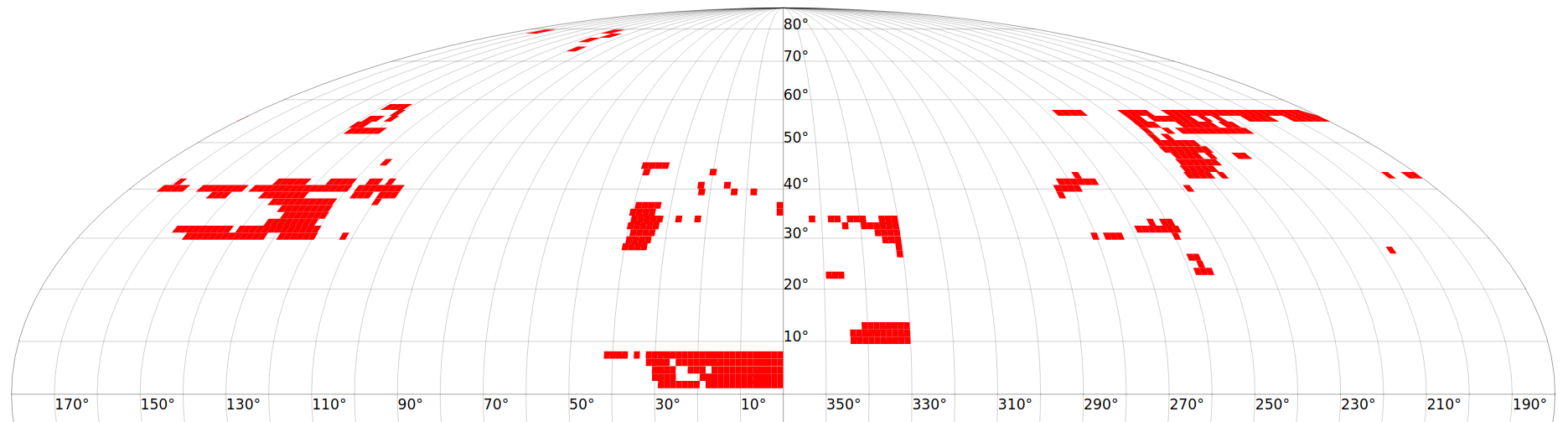}
\caption{J-PLUS DR1 skymap coverage, represented using the International Celestial Reference System (ICRS). Source: J-PLUS website (\protect\url{http://archive.cefca.es/catalogues/jplus-dr1/coverage_map.html}).}
\label{fig:dr1_coverage}
\end{figure*}


J-PLUS observations are performed with the JAST/T80 0.8-m telescope, located in the Observatorio Astron\'omico de Javalambre (OAJ, Teruel, Spain; \citealt{cenarro14}), using the panoramic camera T80Cam \citep{marinfranch15}, with a 2 deg$^2$ field of view (FoV), covering a wavelength region between 0.35 to 1 $\mu$m by means of a 12 filter system: five intermediate-band filters (four of them equivalent to the SDSS \textit{g}', \textit{r}', \textit{i}', and \textit{z}' filters, the other one being a modified version of the SDSS \textit{u}' filter), and seven narrow-band filters designed specifically for J-PLUS (see Fig. \ref{fig:filter_transmissions}). The observational strategy is to obtain, sequentially, three exposures per filter, being the total time of acquisition in the 12 filters for the same field approximately one hour. Taking into account a) the large fraction of sky coverage, b) the spectral resolution, better than previous visible large-area photometric surveys, and c) the observational strategy, we expect to obtain improved spectrophotometry (compared to existing datasets) for quite a large number of minor bodies in the Solar System.

Being such a large survey, the photometric calibration of J-PLUS' data follows a complex procedure. The data presented here relies on a calibration that uses the locus of the Main Sequence stars observed in the footprint of the survey, anchoring the colors to Pan-STARSS's, as described in detail in \citep{LopezSanJuan2019}.

Another interesting feature of the J-PLUS survey is its powerful photometric resolution in the ultraviolet region: five filters are covering the $0.3-0.45$ $\mu$m range. Compared to J-PLUS, the most prominent survey that offers ultraviolet spectral information of minor bodies (for a total of 589 objects) is the Eight Color Asteroid Survey (ECAS, \citealt{ecas}), which has only three filters in the same region: \textit{s}, \textit{u}, and \textit{b}, centered at 0.337, 0.359, and 0.437 $\mu$m respectively (see Fig. \ref{fig:jplus_vs_ecas}). Note that, although the central wavelength of the \textit{b} ECAS filter is 0.437 $\mu$m, this filter covers a broad wavelength range, i.e., it includes non-ultraviolet information, while all the J-PLUS ultraviolet filters are narrow-band (except for the $u$ filter), giving specific coverage for this region. As we will see in the next section, the number of Solar System objects that we expect to recover from J-PLUS observations in this spectral region is approximately one order of magnitude higher than the number of available objects in the literature, which will improve the present knowledge of minor bodies in an spectral region that has not yet been deeply explored.

The present work is organized as follows: in Sect. \ref{section2} we give a brief summary of the survey status, we present the catalog and its most remarkable properties, and we also discuss the outlier removal. In Sect. \ref{section3} we present a method to select the optimal solar colors in the J-PLUS photometric system, needed to later compute the reflectance photospectra. In Sect. \ref{section4} we discuss the data quality and the main issues that need to be addressed. Last, in Section \ref{section5} we summarize the contents of the paper, proposing future analises that should be done, as well as ways to improve the data extraction from the survey observations. 

Finally, we want to stress that the scope of the present work is to select photometric data from minor bodies contained in the first data release (DR1) of J-PLUS, providing a large sample of minor bodies colors, that can be used to compute very-low-resolution spectra (or photospectra), rather than performing in-depth analyses of the obtained data, also noting that the present work is one in a list of articles presenting diverse aspects of J-PLUS DR1 (e.g., \citealt{halphaemission19, lowFEjplus19, ultracooldwarfsjplus19}).


\section{Observations and MOOJa (Moving Objects Observed from Javalambre) catalog}
\label{section2}

Solar System objects move over the sky background. This means that, depending on the exposure time of the image, the proper motion of the object, and the pixel scale of the detector, the object might appear as a trace, instead of a point source, difficulting, or even making impossible the computation of its photometry. Fortunately, the majority of SSOs have proper motions smaller than 1$\arcsec$/min (mainly objects located in the Main Belt and beyond), which, combined with the mean exposure times in every filter (as we will point out in the next paragraph) will be sufficient for most of the asteroids observed by the survey to be detected as point sources, computing their photometry with no issues.

Although the scientific aim of J-PLUS was not specifically oriented to the observation of minor bodies, its observational strategy allows their detection within the survey images. The observations follow a sky tiling: each tile has the size of the T80Cam FoV, and in order to guarantee coverage continuity, neighboring pointings overlap in the FoV limits. For each pointing of the survey, all 12 filters are observed consecutively, following a predefined sequence. Every J-PLUS pointing is made of 36 exposures (3 for each filter), and the 3 exposures in a given filter are consecutive, with a small dithering of $10''$ between them that allows the removal of bad pixels or columns. The three exposures are then combined, resulting in 12 images per field (one per filter). The mean exposure time per image in the broad-band filters is $\sim$1 minute, and in the case of the narrow-band filters, is $\sim$2-3 minutes. The complete filter sequence is the following: $i$, $u$, $J0430$, $J0861$, $J0515$, $J0410$, $r$, $J0378$, $J0660$, $g$, $J0395$, and $z$. This filter acquisition order has been defined in a way that observations between filters with similar central wavelengths are performed within long enough time intervals for non-sidereal movement detection, while, at the same time, minimizing long filter wheel rotations. 

\subsection*{The MOOJa catalog}

The J-PLUS DR1\footnote{\url{http://archive.cefca.es/catalogues/jplus-dr1/}} contains images of 511 fields, collected from November 2015 to January 2018. Its sky coverage is shown Fig. \ref{fig:dr1_coverage}. The total non-overlapped area is $\sim$900 deg$^2$, i.e., approximately $\sim$10\% of the intended area has already been observed. This is the dataset from which the minor bodies data has been extracted. 

The asteroids present in the J-PLUS DR1 are recovered using the \texttt{SSOS} pipeline, a versatile software developed to detect both known and unknown SSOs in astronomical images.\footnote{\url{https://ssos.readthedocs.io/}} A detailed description of the pipeline setup is given in \cite{mahlke19}. \texttt{SSOS} is applied to each J-PLUS field individually. We briefly summarize the pipeline in the following steps: first, all sources present in the 12 images of each field are catalogued using SExtractor \citep{sextractor1996}. Next, the sources detection in the individual images are associated to their corresponding observation epochs using SCAMP \citep{scamp2006}. SSOs are then separated from the stars, galaxies, and other sources within the sample primarily by evaluating their apparent motion. Finally, using the SkyBoT service \citep{berthier06}, the output sample is cross-matched with the known population of asteroids. The astrometry of both known and unknown objects is reported to the Minor Planet Centre (MPC). However, the positions of the unknown objects are not sufficient to determine the orbits on their own. They will be combined with other isolated tracklets by the MPC, hopefully producing orbits of new asteroids. Thus, the unknown objects have been left out of the catalog, since none of them has been confirmed by the MPC as a new discovery. Magnitudes were computed using SExtractor by fitting adaptive apertures to the source's pixels in the image. SExtractor automatically computes instrumental magnitudes from the pixel values inside the aperture by relying on telescope and image metadata present in the FITS headers of the files to calibrate to the AB system.

The majority of the asteroids in the catalog were observed in a single pointing during a complete filter sequence, or run. However, some of these objects were not detected in all the filters, probably due to the target brightness. In addition, there were some cases in which the asteroid was observed in two consecutive runs during the same night (cases in which the run had to be restarted), or even in different nights. Whenever an asteroid was detected in more than one run in the same night, we divided the data into blocks in which the elapsed time from the first to the last filter acquisition is, at maximum, 60 min\footnote{See Section \ref{section4} for details on the 60 min limit.}. Then, we selected the block with the maximum number of different filters, and, when there were more than one block which fulfilled this condition, we selected the block corresponding to the data with the smallest mean errors. If the asteroid was observed in different nights, in only one run each night, we selected the data corresponding to the night with the maximum number of different filters, and again, if this number was the same for the different nights, we selected the data with the smallest errors.

\begin{figure*}
\centering
\includegraphics[width=\hsize]{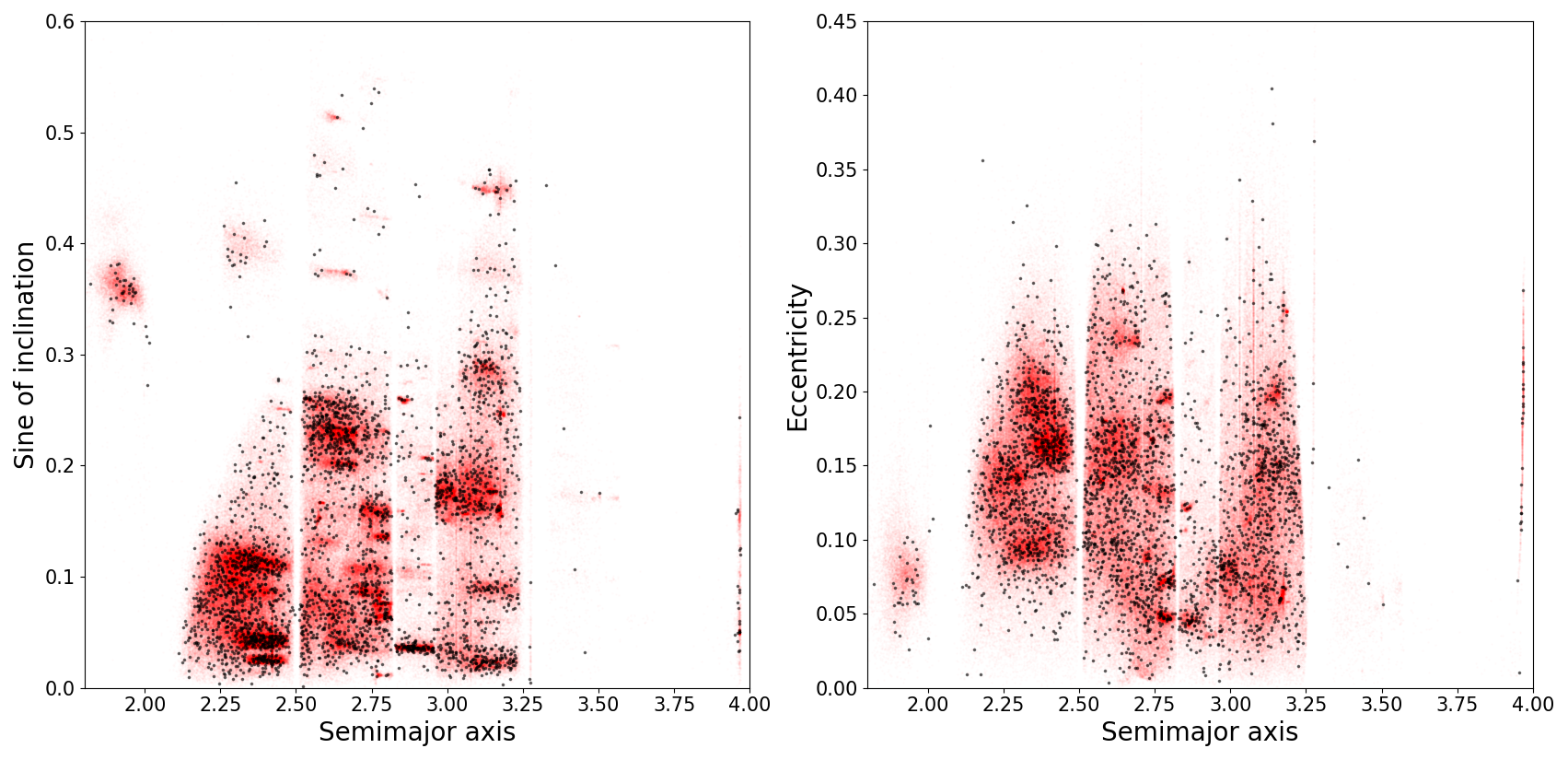}
\caption{Asteroids within the MOOJa catalog which have proper elements available (black) superimposed to all proper elements for minor bodies in the Solar System (red). We represent their semimajor axis values vs the sine of inclination of the orbit (left panel) and vs eccentricity (right panel). Data available at the AstDyS-2 website (\protect\url{https://newton.spacedys.com/~astdys2/propsynth/all.syn}).}
\label{fig:proper_elements}
\end{figure*}

The full catalog presents data for a total of 3\,666 objects: 3\,570 numbered asteroids, plus 96 asteroids with only provisional designation. Out of them, 359 have been observed in all the 12 filters; 386 in, at least, the 5 ultraviolet filters; and 2\,476 in, at least, the 7 non-ultraviolet filters. In Fig. \ref{fig:proper_elements}, we can see the distribution of the observed objects throughout the main asteroid belt.


\begin{figure*}[h!]
\centering
\includegraphics[width=\hsize]{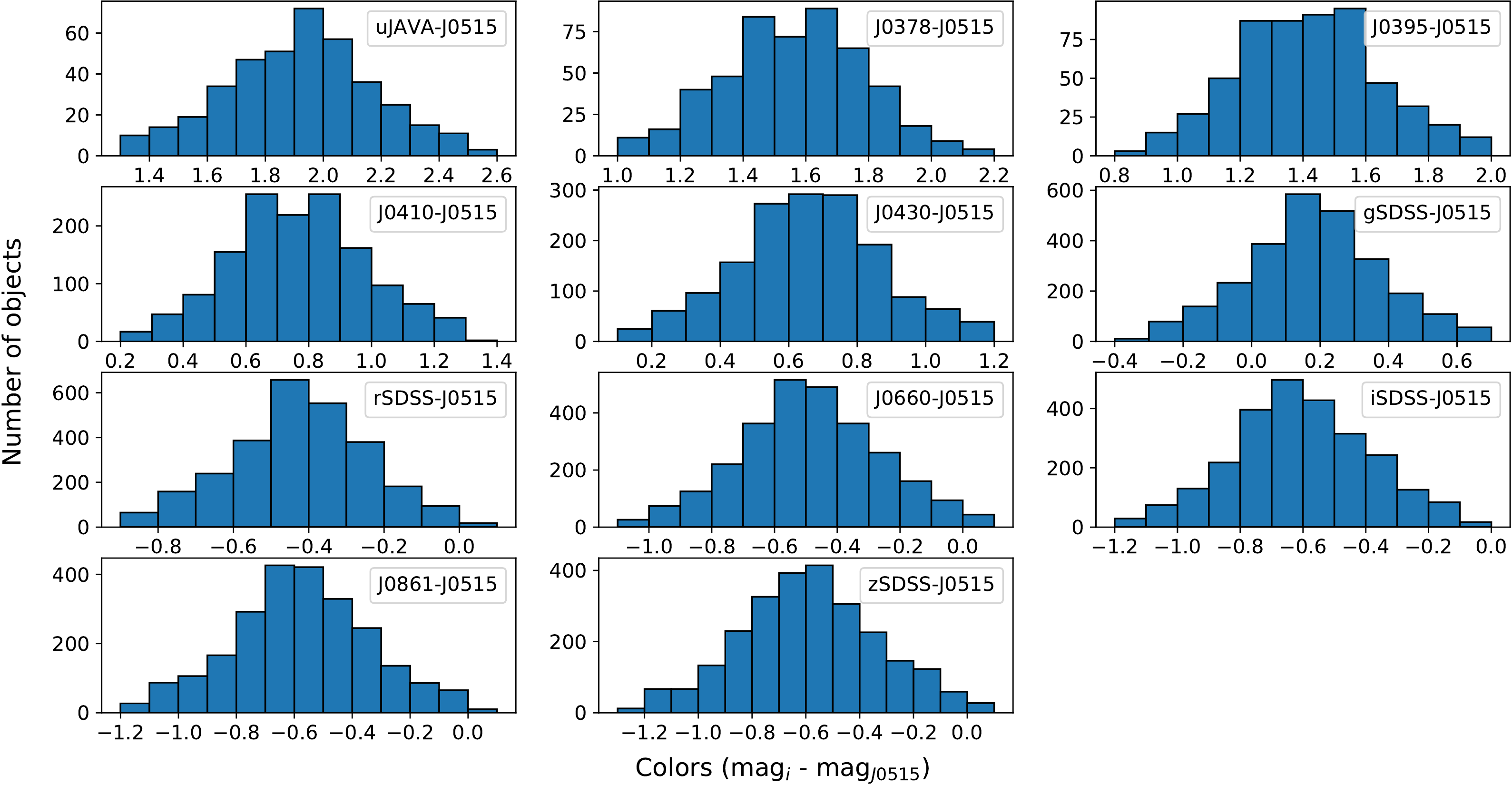}
\caption{Histograms for the color distribution after the sigma clipping was performed, in order to remove outliers. As mentioned in the main text, all the colors are referred to m$_{J0515}$, which is the filter that we have used for normalization. The bin size is 0.10 mags.}
\label{fig:color_histograms}
\end{figure*}

\begin{figure*}
\centering
\includegraphics[width=\hsize]{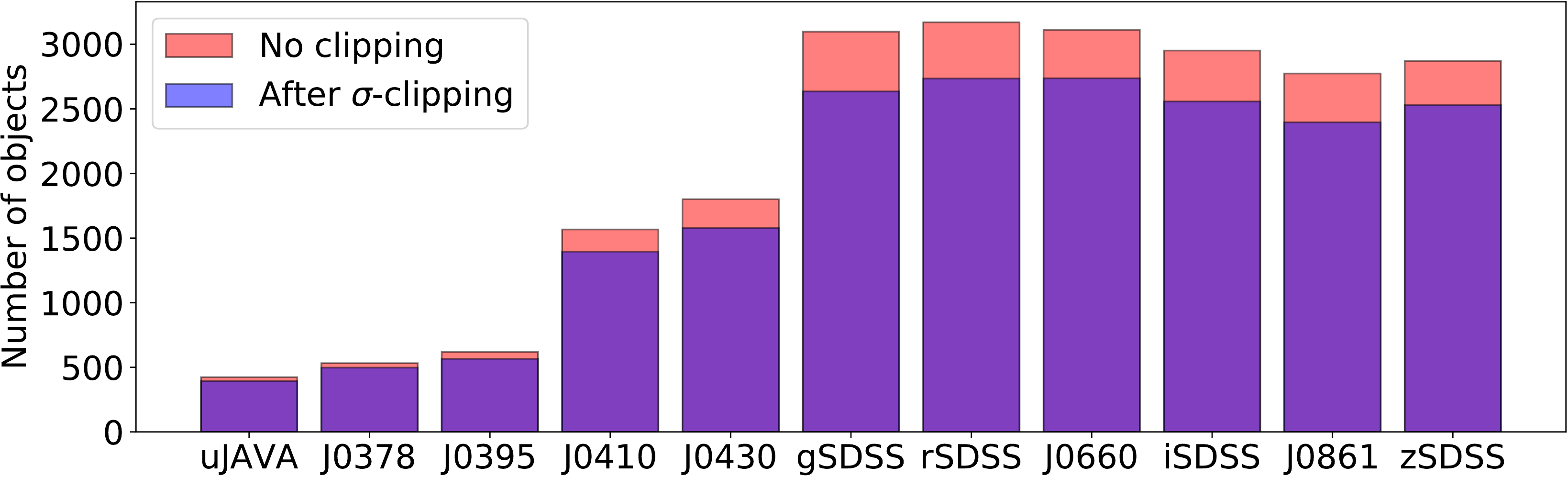}
\caption{Number of objects that have been observed in an specific filter and in J0515.} 
\label{fig:colors_vs_nast_clip_vs_noclip}
\end{figure*}

To remove outliers, we applied a $\sigma$-clipping algorithm in the color space. We do not expect the color distributions to be completely gaussian-like. However, despite this limitation, we proceed assuming that they are, since this assumption does not appear to introduce unwanted features in the data rejection process. We computed colors $m_f-m_{J0515}$\footnote{We chose $m_{J0515}$ as the normalization filter, see below.}, and then we computed the mean and standard deviation of the corresponding distributions. We selected data satisfying $|x-\overline{x}|>2.5\sigma$\footnote{This threshold was established based on a compromise between data completeness and cleanness: more restrictive thresholds removed observations that were not really flawed, while, by being less restrictive, we did not manage to remove data that were obvious outliers}, where $x$ is the color and $\sigma$ is the standard deviation of the sample, discarding the $m_f$ observation of the corresponding color. The process was iterated until no outliers were found. The final color distributions are shown in Fig. \ref{fig:color_histograms}. A total number of 460 objects were removed due to the $\sigma$-clipping procedure. In Fig. \ref{fig:colors_vs_nast_clip_vs_noclip}, we show how many objects have data colors for each filter (referred to J0515), before and after the outlier removal.

We have to keep in mind two things: first, all the objects that do not have observations on the J0515 filter, were removed, since we cannot use this filter to normalize the corresponding photospectrum; second, and related to the previous point, if the observation in the J0515 filter is considered an outlier, that object will also be removed, since, again, even if the other observations are valid, the photospectrum cannot be constructed due to the lack of reliability of the reference observation. Due to this, another 84 entries were removed from the original dataset. Thus, we will provide two catalogs: one, with all the recovered data, in case the reader wants to perform their own analyses, and another, which we consider a robust enough dataset, with the colors referred to the normalization filter ($m_f-m_{J0515}$), ready to compute the corresponding photospectra (see Section \ref{section3}).

\begin{figure}[h!]
\centering
\includegraphics[width=\hsize]{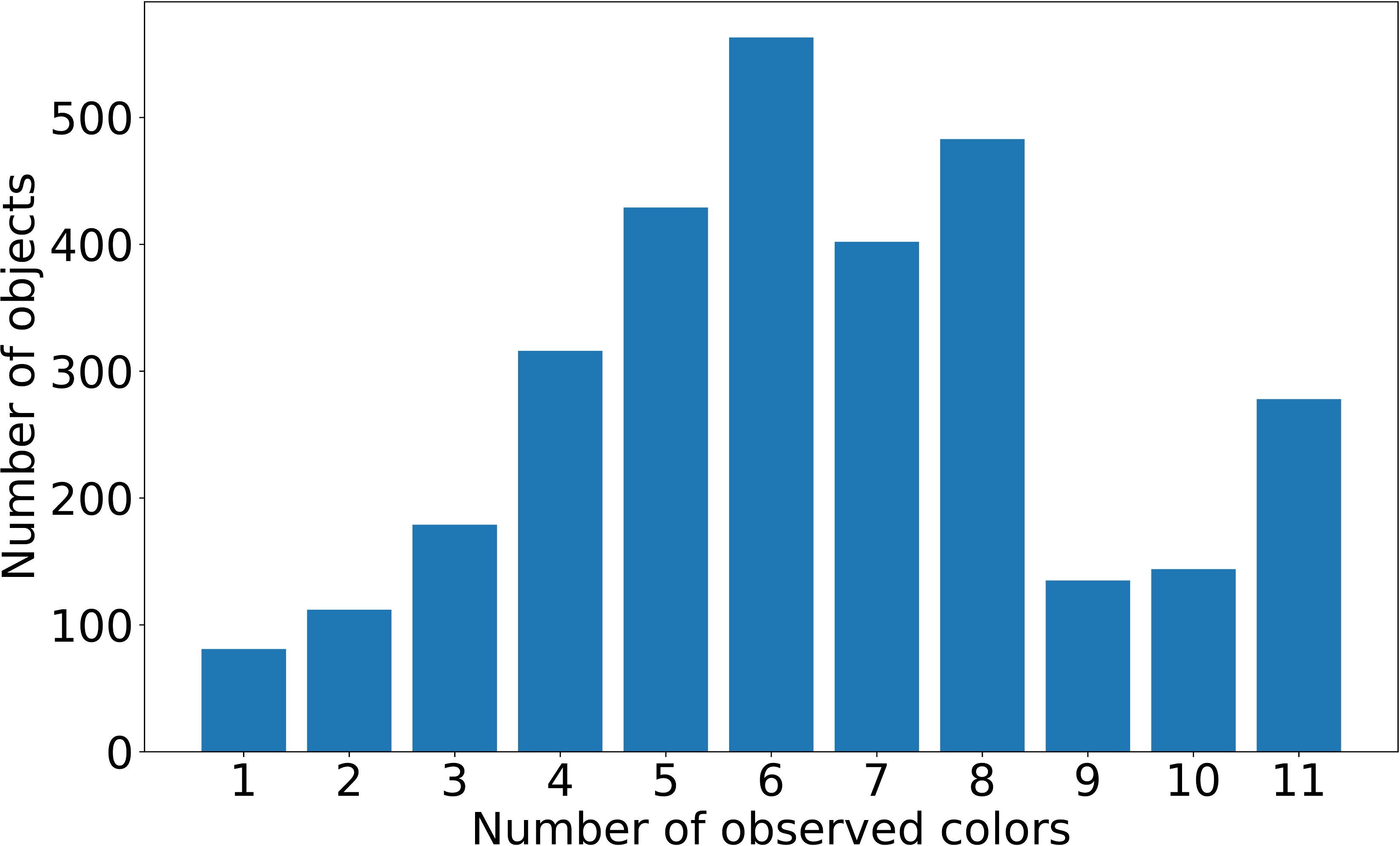}
\caption{Number of objects per quantity of colors remaining after sigma-clipping.}
\label{fig:ncolors_vs_nast}
\end{figure}

The colors catalog provided consists of data for 3122 asteroids. In Appendix \ref{appendix:colorcolor} we show some color-color distributions, as a quick proof-of-concept for the reader (since the main focus of this work is to obtain photospectra rather than doing color analyses) that the data presented in the MOOJa catalog is enough to provide an apparent differentiation between the C and S complexes.
Out of those 3122 objects, 278 have been observed in every filter, thus, we will be able to construct their complete photospectra. In addition, another 2005 objects have data for, at least, 6 colors (i.e., observed in 7 filters, including the normalization one, J0515). In Fig. \ref{fig:ncolors_vs_nast} we show how many objects have data available for a given total number of colors.

\begin{figure*}[htb!]
\centering
\includegraphics[width=\hsize]{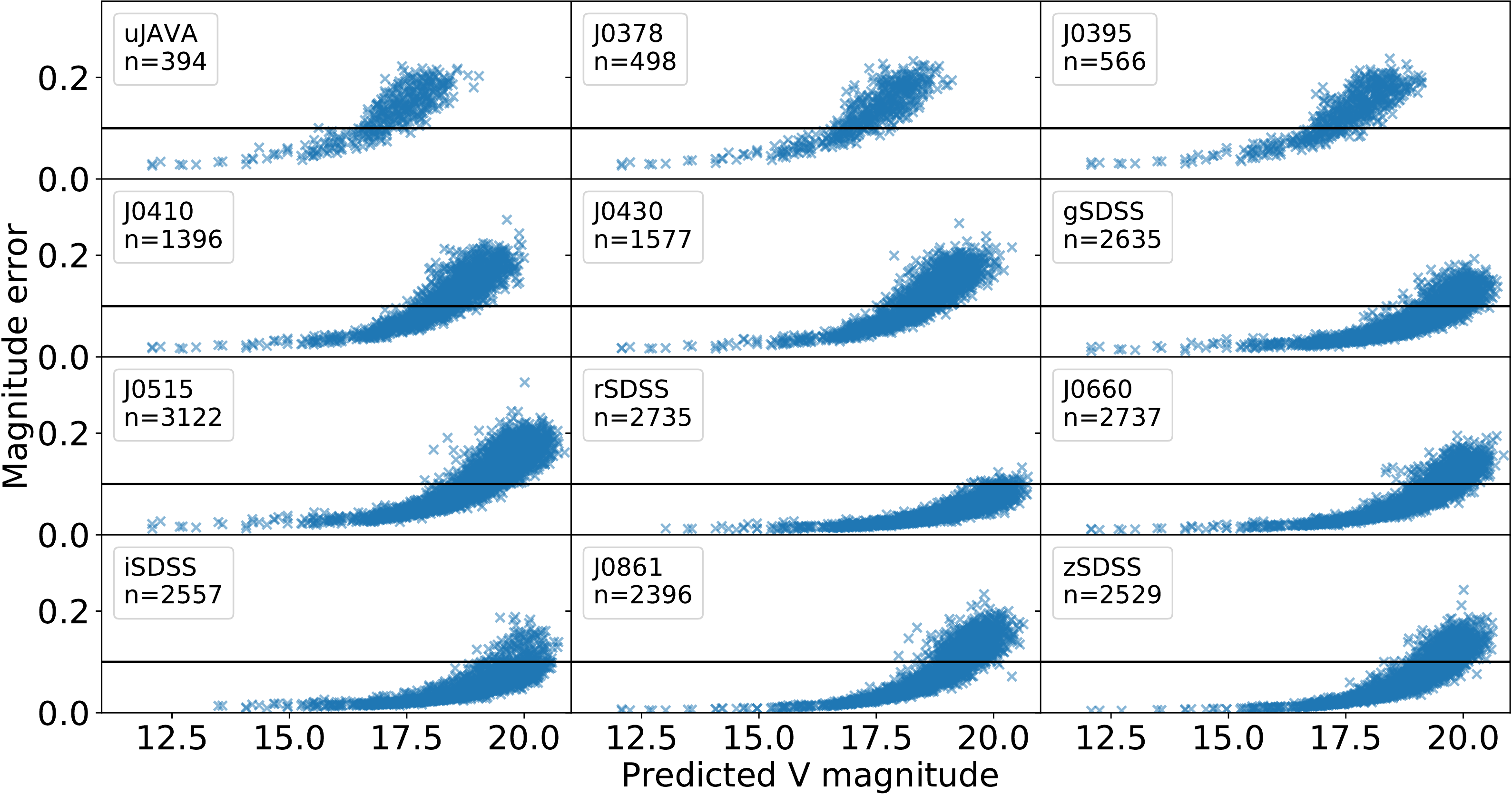}
\caption{Distribution of the errors associated to each observation (for the sigma-clipped dataset), relative to the predicted apparent V magnitude, for the 12 filters of J-PLUS. The horizontal black line corresponds to an error of 0.1 mag.}
\label{fig:vmag_vs_errors}
\end{figure*}

\begin{figure*}
\centering
\includegraphics[width=\hsize]{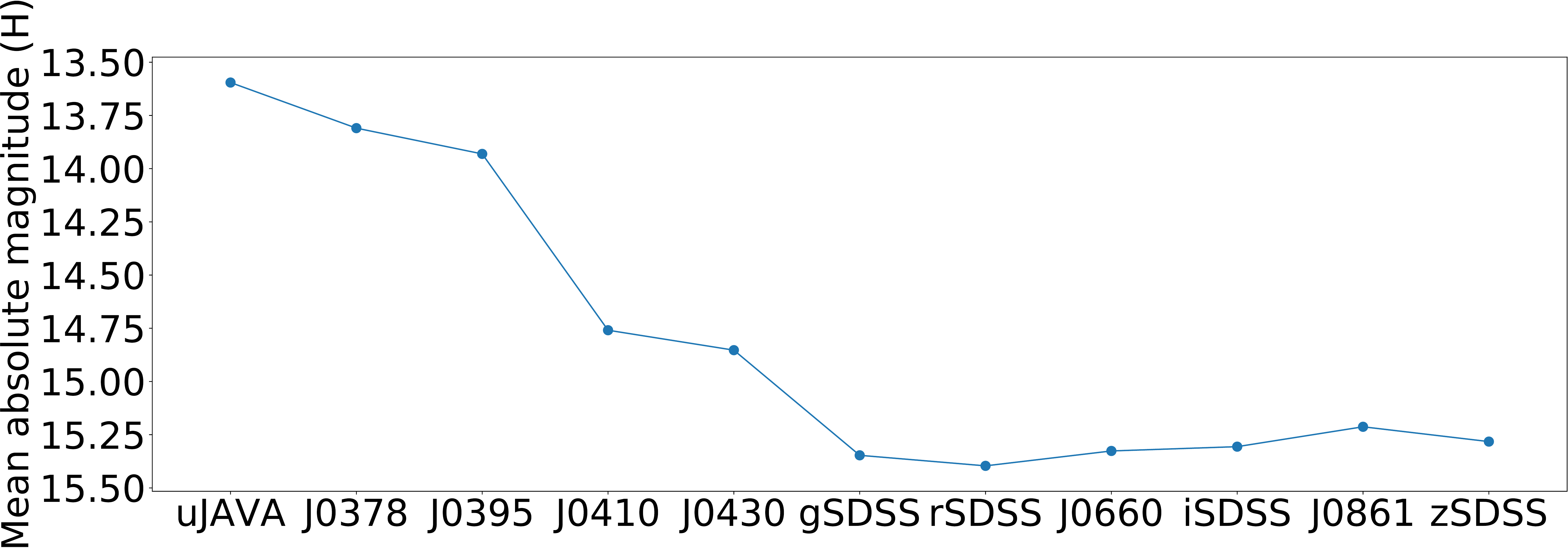}
\caption{Mean absolute magnitudes per color. Magnitudes extracted from the JPL website.}
\label{fig:mean_H_vs_colors}
\end{figure*}

The limiting magnitude of the survey can be inferred by plotting the photometric errors versus the predicted magnitude for the detected objects (see Fig. \ref{fig:vmag_vs_errors}). The limiting magnitude depends on the filter: photometric data with magnitude errors smaller than 0.1 can be obtained, for the uJAVA, J0378, and J0395 filters (UV region), for V magnitudes brighter than V$\sim$17.5; for J0410 and J0430 (transition region from the UV to the visible) this limit goes up to V$\sim$19; finally, for the visible filters (gSDSS, J0515, rSDSS, J0660, iSDSS, J0861, and zSDSS) the limiting magnitude is V$\sim$20.5. This explains why in the redder filters there are data for smaller objects, as opposed to the bluer filters where only those objects which are big enough were detected in the UV region (see Fig. \ref{fig:mean_H_vs_colors}).

Taking into account the area that the DR1 has already covered, a conservative estimate of the total number of minor bodies that will be recovered after the survey is completed is approximately 18\,000 objects, out of which we expect $\sim$1\,800 observed in  all the 12 filters, $\sim$2\,000 in the 5 ultraviolet filters, and $\sim$12\,000 in the 7 non-ultraviolet filters.

The full catalog consists of a list of asteroids, labeled by their identification number or provisional designation (first column of the catalog). After the ID number, there are three columns regarding each of the 12 filters: the first column is the detected magnitude, the second is the error in the magnitude measurement, and the third is the MJD at the middle of the three exposures in the corresponding filter. Each asteroid has also associated (in the case it has been computed) their corresponding taxonomic classification, according to \cite{carvano2010}, computed using SDSS data, in the second-to-last column of the catalog. The absolute magnitudes, H (from the JPL website), for each object are in the next-to-last column. Finally, in the last column, and for quick reference, it is shown the number of filters in which the object has been observed.

The colors catalog presents the colors, as $m_f-m_{J0515}$. The catalog also has the identification number of each entry in the first column. Then, it provides 4 entries per color, with the color value, its corresponding error, the elapsed time between the observation in both filters, and the MJD of the observation in the $m_f$ filter. As it was the case for the full catalog, the columns with taxonomy and  absolute magnitude are included. In this case, we also provide the predicted V magnitude at the moment of the observation (extracted using the JPL Horizons service) in the next-to-last column. The last column shows the number of colors for which each object has data available (analogous to the last column of the full catalog).

Both datasets can be downloaded from the CDS repository [link will be inserted here once the paper is accepted.].


\section{The solar colors in the J-PLUS photometric system}
\label{section3}

Solar System objects reflect the light of the Sun at the wavelengths observed by J-PLUS. It is, therefore, necessary to remove this signature to obtain reflectance values. These reflectance values, the spectra of minor bodies, are used to infer their surface composition, by comparison with meteoritic samples.

In order to transform J-PLUS magnitudes into reflectances, we have followed a simple equation, where the reflectance value for each filter, $R_f$, is computed as
\begin{equation}
\label{eq:reflectance}
R_f  = 10^{-0.4(C_{f,ast}-C_{f,\odot})}
\end{equation}
where $C_f=m_f-m_{J0515}$ is the color, or magnitude diference, between filter $f$ and the normalization filter ($J0515$), for both the asteroid and the Sun, at the nominal central wavelengths for each filter (see Table \ref{table:jplusfilters}).


\begin{table}
\caption{Central wavelengths and bandwidths of the J-PLUS filters. Source: \protect\url{http://www.j-plus.es/survey/instrumentation}.} 
\label{table:jplusfilters} 
\centering 
\begin{tabular}{l l c c} 
\hline\hline 
ID & Filter Name & Central $\lambda$ ($\mu$m) & Bandwidth ($\mu$m) \\ 
\hline 
1  & $u$ & 0.3485 &  0.0508 \\ 
2  & $J0378$ & 0.3785 &  0.0168 \\
3  & $J0395$ & 0.3950 &  0.0100 \\
4  & $J0410$ & 0.4100 &  0.0200 \\
5  & $J0430$ & 0.4300 &  0.0200 \\
6  & $g$ & 0.4803 &  0.1409 \\
7  & $J0515$ & 0.5150 &  0.0200 \\
8  & $r$ & 0.6254 &  0.1388 \\
9  & $J0660$ & 0.6600 &  0.0145 \\
10 & $i$ & 0.7668 &  0.1535 \\
11 & $J0861$ & 0.8610 &  0.0400 \\
12 & $z$ & 0.9114 &  0.1409 \\
\hline 
\end{tabular}
\end{table}

From Eq. \ref{eq:reflectance} it is clear that solar colors ($C_{f,\odot}$) in the pertinent filters are needed to obtain reflectances. Values of $C_{f,\odot}$ are not yet determined by the collaboration, at least not for all filters involved. Therefore, in order to obtain reliable $C_{f,\odot}$ we have used nine sets of data, following two different approaches: one theoretical (four sets) based on published spectra of the Sun, and one empirical (five sets) based on J-PLUS colors of solar type stars.

In the next subsections we provide a more detailed description of the solar spectra that we used, as well as the solar analog selection process. In the last subsection, we will discuss which one would be the best solar colors choice, and how do we select it.

\subsection{Theoretical solar colors}

There exist several solar spectra in the literature that have been and are used by the community. These are all extraterrestrial solar irradiance spectra, obtained at airmass zero, and are all based on data from satellites, space shuttle missions, high-altitude aircrafts, rocket soundings, ground-based solar telescopes, and (or) modeled spectral irradiance. Out of all the available spectra, we arbitrarily chose four of them, in order to have enough different sources and results. The ones used for this analysis were:
\begin{itemize}
\item 1985 Wehrli Standard Extraterrestrial Solar Irradiance Spectrum
\item 2000 ASTM Standard Extraterrestrial Spectrum Reference E-490-00
\item MODTRAN ETR Thuillier Spectrum
\item PMOD/WRC Solar Reference Spectrum 
\end{itemize}
Further references can be found in \cite{wehrli85,frolich97,thuillier2003,thuillier2004,chancekurucz2010}, and \cite{haberreiter17}. All four spectra are available online\footnote{The first three spectra are available at \url{https://rredc.nrel.gov/solar//spectra/}, and the PMOD spectrum is available at the public FTP of the PMOD/WRC (\url{ftp://ftp.pmodwrc.ch/pub/data/SolarReferenceSpectrum/}).}. To compute $C_{f,\odot}$, we convolved the four spectra with the J-PLUS filters transmission. The detector efficiency and atmospheric transmission information are already included in the filter transmission curves. The computed solar colors are shown in Table \ref{tab:solar_colors}.

\subsection{Empirical solar colors}

The second approach to solve the problem of obtaining the solar colors in the J-PLUS photometric system was to analyze the DR1 searching for a set of possible Sun-like stars observed in the 12 filters. Solar-type stars, solar analogs, and solar twins are stars that are particularly similar to the Sun, the solar twin being most like the Sun followed by solar analogs and then solar-types \citep{soderblom98}. 
Solar twins should be indistinguishable from the Sun, this means:
\begin{enumerate}
\item Stellar parameters similar to those of the Sun
\item Age within $\sim$1 Gyr, so that the evolutionary state is comparable.
\item No known stellar companion, because the Sun has none.
\end{enumerate}
We stress here that this type of stars are not particularly targeted by the survey strategy and we have to rely on serendipitous observations. The most commonly used stellar parameters are effective temperature, gravity, and metallicity (represented as $T_{\textrm{eff}}$, $\log g$, and [Fe/H], respectively). In the Sun's case, these parameters are $T_{\textrm{eff}}=5777\pm10$K, $\log g=4.4374\pm0.0005$, and zero metallicity\footnote{The metallicity of a star is defined relative to the Sun.}, $[\textrm{Fe}/\textrm{H}]=0$ (see \citealt{smalley05} and references within).

{Stellar parameters can be computed from J-PLUS data. We used a supervised machine learning model discussed by Galarza et al. (in prep.), based on the random forest algorithm (see \citealt{breiman2001} and references within) to compute them. }

The model needs a subset of well-known data in order to be trained: this training subset was obtained by cross-checking the J-PLUS DR1 catalog with the SEGUE survey from the SDSS. It uses a set of multiple J-PLUS colors for a given object as input parameters, and then returns estimated values of effective temperature, surface gravity, and metallicity as output parameters.

Using this method, we managed to select a list of 8633 stars within J-PLUS, with $5767\textrm{K}<T_{\textrm{eff}}<5787\textrm{K}$, thus meeting the first criterion: we will refer to this dataset as the G1 set. The remaining criteria, however, were not fully achievable due to the low spectral resolution and the need of a thorough analysis of the data in the G1 set, which is out of the scope of the present work. To compensate for this, we selected several subsets of the G1, producing a total of five sets of Sun-like stars, that we named G1--G5. These subsets were arbitrarily selected, filtered using the following criteria (the number of objects in each group is shown between parentheses):

\begin{itemize}
\item G1: Stars with $5767\textrm{K}<T_{\textrm{eff}}<5787\textrm{K}$ (8633)
\item G2: Stars with $5772\textrm{K}<T_{\textrm{eff}}<5782\textrm{K}$ (4792)
\item G3: Stars with $5767\textrm{K}<T_{\textrm{eff}}<5787\textrm{K}$ and photometric errors in every filter less than 0.01 magnitudes (691)
\item G4: Stars with $5772\textrm{K}<T_{\textrm{eff}}<5782\textrm{K}$ and photometric errors in every filter less than 0.01 magnitudes (409)
\item G5: Stars with $5767\textrm{K}<T_{\textrm{eff}}<5787\textrm{K}$, $-0.3<[\textrm{Fe}/\textrm{H}]<0.3$ and $4.1<\log g<4.7$ (71)
\end{itemize}

In Table \ref{tab:solar_colors} we can see $C_{f,\odot}$ derived for the G1 to G5 datasets.


\begin{table*}
\caption{Solar colors for each J-PLUS filter, computed from solar spectra (theoretical) and from Sun-like stars observed in J-PLUS in all 12 filters (empirical). All the colors are referred to the $J0515$ filter, the normalization filter that we have used for J-PLUS spectrophotometric data (i.e., m$_i$ - m$_{J0515}$).} 
\label{tab:solar_colors} 
\centering 
\begin{tabular}{l c c c c c c c c c c c} 
\hline\hline 
Spectrum/ & $u$ & $J0378$ & $J0395$ & $J0410$ & $J0430$ & $g$ & $r$ & $J0660$ & $i$ & $J0861$ & $z$\\ 
Dataset & & & & & & & & & & & \\
\hline 
\textit{Theoretical} & & & & & & & & & & & \\
\hline 
1985 Wehrli & 1.4913 & 1.2941 & 1.1518 & 0.5876 & 0.5118 & 0.1757 & -0.2877 & -0.3269 & -0.3979 & -0.4180 & -0.4202\\
2000 ASTM & 1.5110 & 1.3023 & 1.1455 & 0.5853 & 0.5097 & 0.1750 & -0.2879 & -0.3275 & -0.3985 & -0.4173 & -0.4338\\
Thuillier & 1.4325 & 1.2426 & 1.1240 & 0.5561 & 0.4993 & 0.1614 & -0.2777 & -0.3157 & -0.4048 & -0.4387 & -0.4251\\
PMOD/WRC & 1.4749 & 1.2665 & 1.1499 & 0.5707 & 0.4913 & 0.1605 & -0.2874 & -0.3156 & -0.4072 & -0.4154 & -0.4245\\
\hline 
\textit{Empirical} & & & & & & & & & & & \\
\hline
G1 & 1.4048 & 1.1210 & 1.0886 & 0.5541 & 0.5015 & 0.1456 & -0.3288 & -0.3752 & -0.4997 & -0.5330 & -0.5587\\
G2 & 1.4042 & 1.1195 & 1.0883 & 0.5537 & 0.5013 & 0.1455 & -0.3288 & -0.3753 & -0.4994 & -0.5328 & -0.5585\\
G3 & 1.4828 & 1.2037 & 1.1375 & 0.5778 & 0.5125 & 0.1489 & -0.3265 & -0.3742 & -0.4903 & -0.5169 & -0.5443\\
G4 & 1.4808 & 1.2014 & 1.1361 & 0.5768 & 0.5125 & 0.1483 & -0.3261 & -0.3746 & -0.4901 & -0.5169 & -0.5444\\
G5 & 1.5256 & 1.2784 & 1.1909 & 0.5853 & 0.5123 & 0.1579 & -0.3296 & -0.3728 & -0.4923 & -0.5167 & -0.5331\\
\hline 
\end{tabular}
\end{table*}

\subsection{The optimal solar colors for MOOJa photospectra}

In order to check for the quality of the determined solar colors and to select the dataset that better reproduces the asteroid reflectances, we used a sample of known spectra of minor bodies, and then compared it with the reflectances that we retrieved from the MOOJa catalog using each set of solar colors.

The spectra of minor bodies were taken from the several dedicated surveys that have observed, up to now, more than 3\,000 spectra in the visible spectral range. The most prominent ones, due to the number of observed objects, as well as their spectral resolution, are SMASS (I and II, \citealt{smassI,smass_observations}), S3OS2 \citep{s3os2}, and PRIMASS \citep{deleon2018_primassL}. We have used these, together with other well-known datasets (ECAS, \citealt{ecas}; Sawyer, \citealt{sawyer2005}; 24 Color-Asteroid-Survey, \citealt{24cas}) for comparison with the J-PLUS spectrophotometry.

For each object present in the literature and in J-PLUS DR1 we first convolved their spectra with the J-PLUS transmission filters to obtain the reflectances in each fitler using
\begin{equation}
\label{eq:convolution}
R_f  = \frac{\sum\limits_{\lambda} T_f S \lambda d\lambda}{\sum\limits_{\lambda} T_f \lambda d\lambda},
\end{equation}
where $T_f$ is the filter transmission, $S$ is the spectrum reflectance, and $\lambda$ and $d\lambda$ are, the wavelength of the spectrum and its step, respectively. This means that, to convolve the spectra with the filters, we needed to interpolate the filter transmissions at the corresponding spectral wavelengths, in order to have one point of the filter transmission curve for each reflectance and wavelength value.

In the cases where the error information associated to each point of the literature spectra was available, the errors were propagated according to uncertainty propagation theory (this is the case of SMASSII, S3OS2, ECAS, and 24CAS). When the spectra did not have error information available (SMASSI, Sawyer, and PRIMASS), the errors were computed as the absolute value of the difference, point to point, between the spectrum and its fourth-order polynomial fit. The corresponding photospectra, together with the J-PLUS observations, are shown in Appendix \ref{appendix:jplus_vs_literature}.

There are a number of factors that may account for differences between spectra observed by different surveys, ranging from actual spectral variation in the asteroids to survey systematic factors and choices of solar analog stars. Thus, we decided to parameterize this difference, taking into account the differences in reflectance, as well as the errors associated to both spectrum and photospectrum. This parameter, $Q$, roughly quantifies how similar is one literature spectrum to its corresponding J-PLUS photospectrum. Its definition is:
\begin{equation}
\label{eq:quality}
\delta Q_i = \sqrt{\frac{(R_{i,Lit.}-R_{i,JPLUS})^2}{\sigma_{i,Lit.}^2+\sigma_{i,JPLUS}^2}},
\\
Q = \frac{\sum\limits_{i=1}^{N} \delta Q_i}{N-1},
\end{equation}
where $R_{i,Lit.}$ is the reflectance value for the filter $i$ obtained after the convolution of the literature spectrum with the J-PLUS filters, $R_{i,JPLUS}$ is the reflectance value for the filter $i$ computed from the J-PLUS magnitudes after removing the solar colors; $\sigma_{i,Lit.}$ and $\sigma_{i,JPLUS}$ are their respective errors; and finally, $N$ is the number of filters in which the asteroid has been observed in J-PLUS.

Each J-PLUS photospectrum has as many associated $Q$ indexes as available spectra in the literature (see Table \ref{tab:jplus_vs_literature} for information of the analyzed photospectra). Also, for each solar colors set, we thus have a $Q$ distribution. Due to the definition of the $Q$ index, we expect that, the closer the value of $Q$ is to zero, the more similar a J-PLUS spectrum is to its literature counterparts. Thus, we will consider the distribution with the smallest mean value of $Q$ as the best approximation, and we will select the solar colors that produced that distribution as the best suited to compute the photospectra of the asteroids in the MOOJa catalog. The analysis of the resulting distributions is discussed in the next paragraphs.


\begin{table}
\caption{List of asteroids within the MOOJa catalog that have published spectra or photospectra in different databases. The first column indicates the identification number of the asteroid, and the second column shows the survey(s) where the asteroid was observed. The numbers correspond to: 1-ECAS, 2-24CAS, 3-SMASSI, 4-SMASSII, 5-S3OS2. 6-Sawyer, 7-PRIMASS. The third and fourth columns show, respectively, the number of J-PLUS filters in which the asteroid has been observed, and the time elapsed (in minutes) from the exposure in the first filter to the exposure in the last one. Columns five and six are the date and the time at mid-observation.} 
\label{tab:jplus_vs_literature} 
\centering 
\begin{tabular}{l l c c c c} 
\hline 
Asteroid & Survey & N$_f$ & $t_{\rm{Total}}$ & Date & Time\\ 
\hline\hline
90     &  1,2,4,5  & 10 & 35.2 & 2017-10-14 & 22:31:33\\
95     &  1,4,5,6  & 10 & 35.3 & 2017-10-13 & 21:15:57\\
122    &  2,4      & 10 & 36.0 & 2017-09-14 & 01:00:37\\
159    &  1,4      & 10 & 31.3 & 2017-11-21 & 21:34:23\\
184    &  4,5      & 12 & 38.0 & 2017-09-21 & 02:55:42\\
277    &  1        & 11 & 75.2 & 2017-08-14 & 02:19:06\\
413    &  2,4      &  9 & 36.3 & 2017-03-29 & 20:08:07\\
462    &  2,4      & 12 & 45.1 & 2017-11-26 & 20:10:27\\
712    &  1,2,4,6  &  9 & 31.2 & 2017-10-15 & 18:55:39\\
758    &  2        &  9 & 31.2 & 2017-11-21 & 23:13:51\\
784    &  4        & 12 & 39.4 & 2016-12-30 & 02:32:55\\
851    &  1,3      & 10 & 38.9 & 2017-09-28 & 22:13:18\\
1024   &  4,5      & 12 & 64.2 & 2017-01-08 & 01:43:09\\
1212   &  1,2,4    & 12 & 37.9 & 2018-01-16 & 19:24:56\\
1245   &  1        & 12 & 38.0 & 2017-11-19 & 23:51:23\\
1677   &  5        & 12 & 43.9 & 2017-10-29 & 02:19:01\\
1796   &  1,4,5    & 12 & 39.9 & 2017-08-22 & 01:47:24\\
1923   &  4,7      & 12 & 37.9 & 2017-09-20 & 00:00:53\\
2149   &  3        & 12 & 38.8 & 2017-02-20 & 23:08:27\\
2251   &  4        & 12 & 39.9 & 2017-08-22 & 02:57:29\\
2645   &  3        & 12 & 41.4 & 2017-10-27 & 04:20:00\\
2730   &  4        & 12 & 38.1 & 2017-09-21 & 00:10:14\\
2902   &  4        & 12 & 37.9 & 2017-09-19 & 23:11:23\\
3259   &  5        & 12 & 41.1 & 2017-10-13 & 18:48:10\\
3885   &  4        & 12 & 38.2 & 2017-11-21 & 22:22:45\\
4733   &  4        & 12 & 39.7 & 2017-02-17 & 23:55:11\\
4993   &  4        & 12 & 38.2 & 2017-10-17 & 00:49:08\\
6661   &  7        & 12 & 50.5 & 2017-08-15 & 03:09:08\\
6769   &  7        & 12 & 38.1 & 2017-10-15 & 00:04:47\\
7274   &  7        &  9 & 36.9 & 2016-12-07 & 02:16:50\\
34339  &  7        & 10 & 38.0 & 2017-09-24 & 03:18:31\\
85167  &  7        &  9 & 30.1 & 2017-01-07 & 00:43:29\\
113374 &  7        &  8 & 38.1 & 2017-09-23 & 22:48:49\\

\hline
\end{tabular}
\end{table}

\begin{figure}
\centering
\includegraphics[width=\hsize]{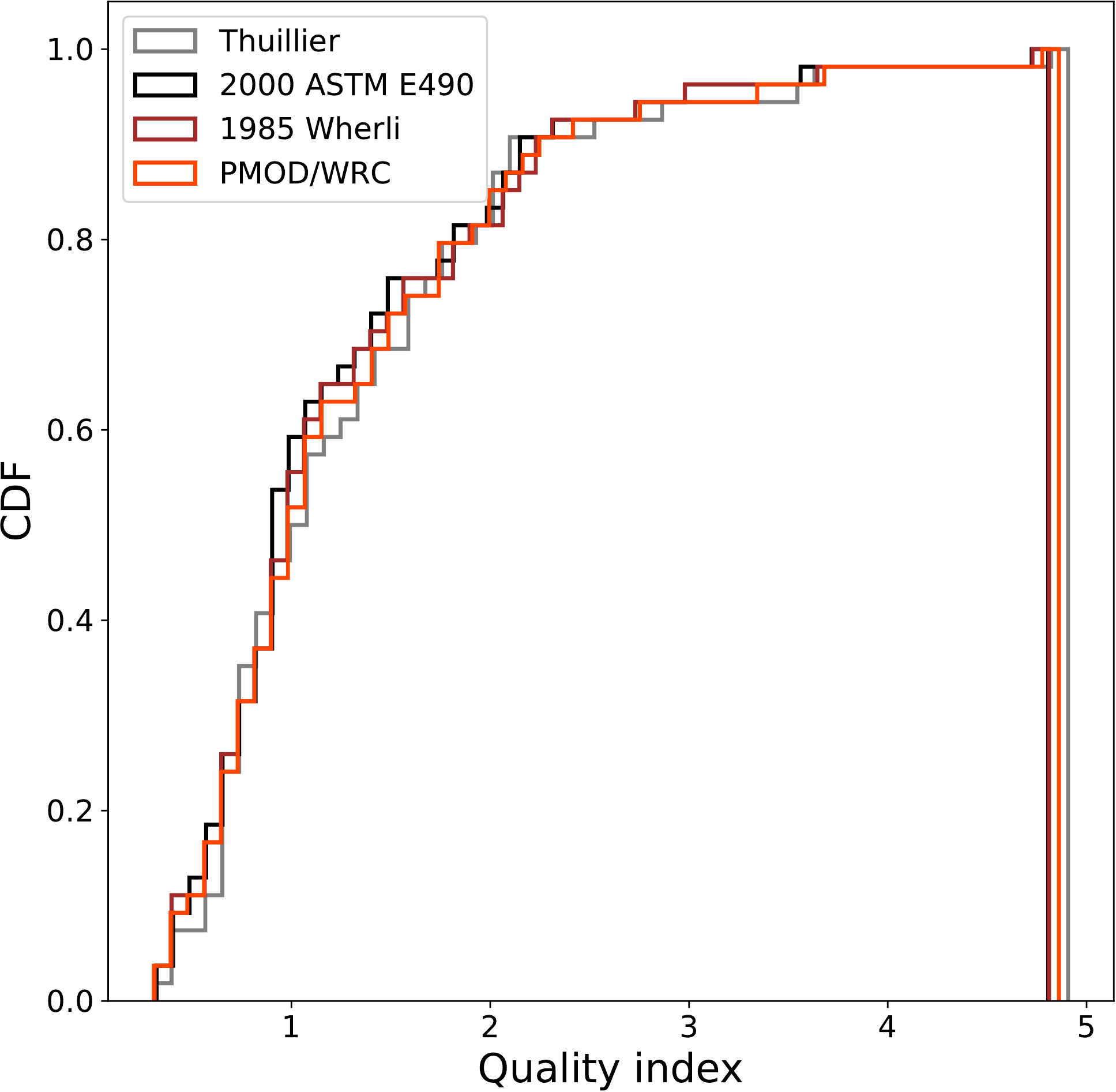}
\caption{Empirical cumulative distribution functions of the $Q$ index obtained after the comparison of the J-PLUS spectrophotometric data with the literature spectra, for the four different solar colors choice.}
\label{fig:cdfs_th_solarcolors}
\end{figure}

The solar colors shown in Table \ref{tab:solar_colors} were used to construct the photospectra of the asteroids in Table \ref{tab:jplus_vs_literature}. Then, after comparing these with the literature spectra, we obtained the $Q$ distribution associated to the use of each solar colors set. In Fig. \ref{fig:cdfs_th_solarcolors} we show the cumulative distribution functions (CDF) for every $C_{f,\odot}$ set derived from a different solar spectrum. The mean values of $Q$ for the four distributions are shown in Table \ref{tab:meanQ}. We find that the CDF with the minimum mean value of $Q$ is the one associated to the 2000 ASTM E-490 spectrum; thus, we selected the corresponding solar colors as the best choice derived from the theoretical approach.


\begin{figure}
\centering
\includegraphics[width=\hsize]{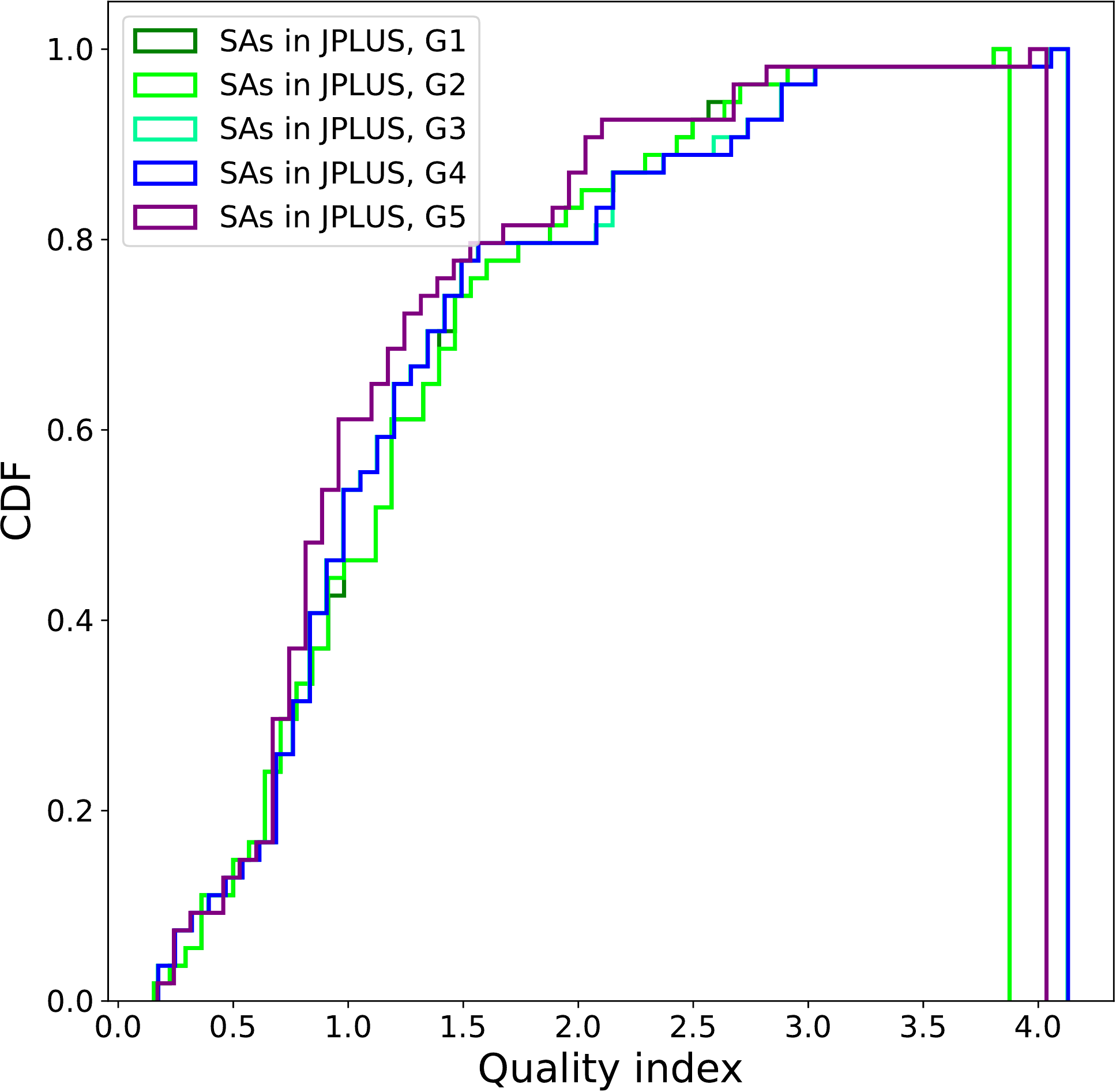}
\caption{Empirical cumulative distribution functions of the $Q$ index obtained after the comparison of the J-PLUS spectrophotometric data with the literature spectra, for the five different solar colors choice.}
\label{fig:cdfs_em_solarcolors}
\end{figure}

In the case of the empirical colors, we note that the colors produced by sets G1 and G2 are very similar, as well as those related to G3 and G4. This is confirmed by the CDFs (shown in Fig. \ref{fig:cdfs_em_solarcolors}) of the $Q$ index associated to the use of these sets of $C_{f,\odot}$, where the dark green and cyan curves are mainly overlapped by the light green and blue distributions, respectively. The mean values of the $Q$ distribution for these cases are shown in Table \ref{tab:meanQ}. The smallest mean corresponds to the solar colors derived from the G5 set, something that, in principle, we might expect, since the stars contained in this group are, according to their stellar parameters, the most similar to the Sun.


\begin{table}
\caption{Mean values and their corresponding sigmas for the $Q$ distributions derived from the use of solar colors computed from Sun spectra (theoretical) and also from the ones obtained after computing the mean colors of different datasets of J-PLUS Sun-like stars (empirical).} 
\label{tab:meanQ} 
\centering 
\begin{tabular}{l c c} 
\hline 
Spectrum/ & &  \\ 
Dataset & $\overline{Q}$ & $\sigma_Q$\\
\hline
\textit{Theoretical} & & \\
\hline 
1985 Wehrli     & 1.31 &  0.81 \\ 
2000 ASTM E-490 & 1.27 &  0.80 \\
Thuillier       & 1.36 &  0.83 \\
PMOD/WRC        & 1.33 &  0.83 \\
\hline 
\textit{Empirical} & & \\
\hline 
G1 & 1.27 &  0.77 \\ 
G2 & 1.28 &  0.77 \\
G3 & 1.28 &  0.82 \\
G4 & 1.28 &  0.82 \\
G5 & 1.16 &  0.73 \\
\hline
\end{tabular}
\end{table}


Taking into account the previous discussion, we adopt $C_{f,\odot}$ derived from the stars in the G5 group as our solar colors because this is the set that produces the smallest $Q$ index mean, i.e., the smallest differences between the photospectra in J-PLUS and their spectroscopy counterparts in the literature. In Fig. \ref{fig:jplus_obs_12filters} (see Appendix \ref{appendix:jplusphotospectracollection}) we show a sample of 74 photospectra of asteroids in the MOOJa catalog, computed using our adopted colors. The quality of the photospectra and the improved resolution with respect to previous photometric surveys will let us perform taxonomical classifications for a large number of minor bodies.

However, we note that there exists a systematic reddening in the visible region of the resulting reflectance photospectra when using the theoretical solar colors. Although this has been investigated, we did not find any satisfactory explanation for this effect. The mean reflectance variations that are induced, depending on which solar colors are used to compute the photospectra (2000 ASTM E-490 as the best theoretical choice, G5 stellar group as the empirical equivalent) are as follows: gSDSS, 0.0153$\pm$0.0013; rSDSS, 0.042$\pm$0.003; J0660, 0.047$\pm$0.005; iSDSS, 0.099$\pm$0.012; J0861, 0.102$\pm$0.011; zSDSS, 0.101$\pm$0.015 (all in magnitude units).

\begin{figure}[!h]
\centering
\includegraphics[width=\hsize]{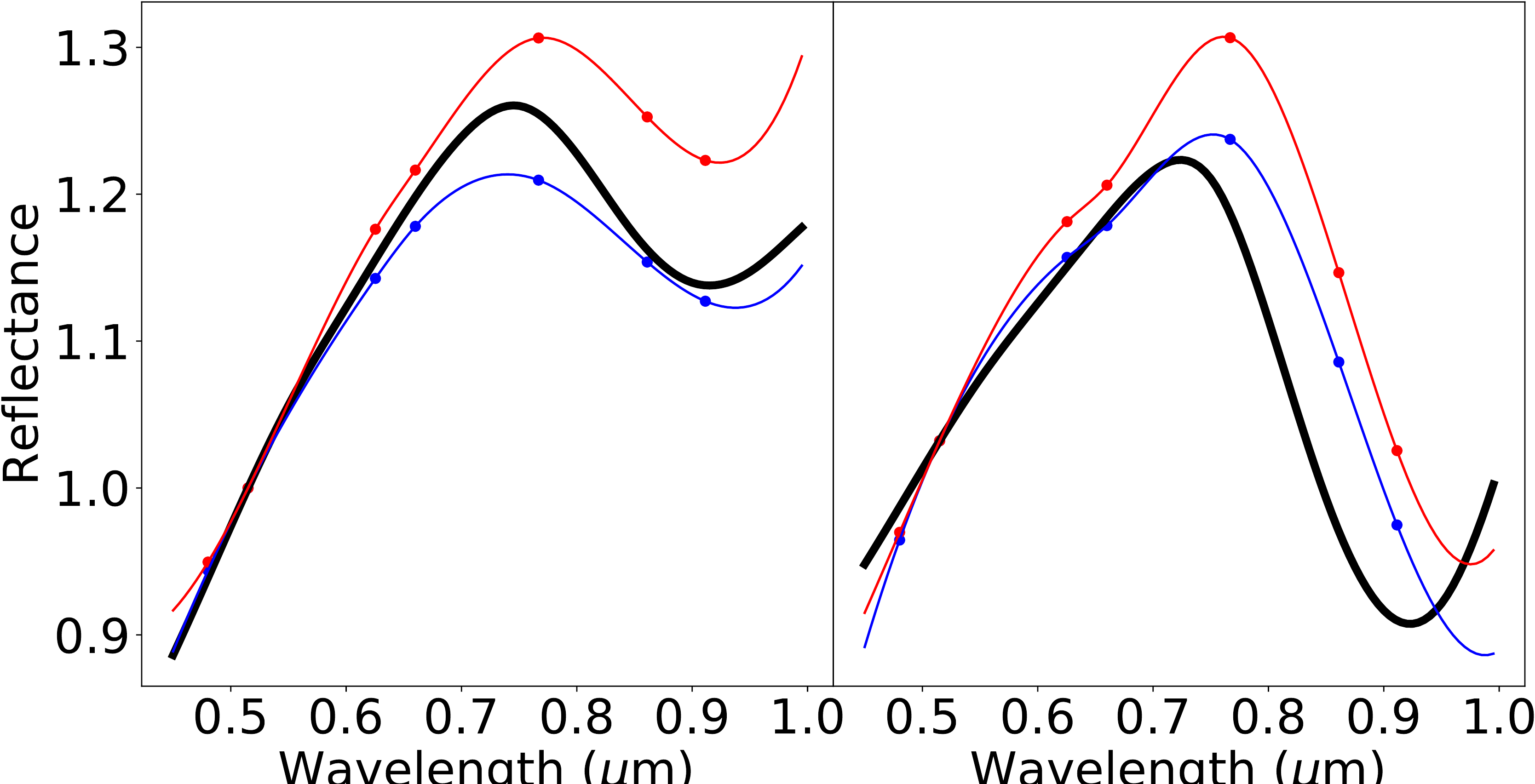}
\caption{Mean photospectra for a subset of S-types (left) and V-types (right) from Appendix \ref{appendix:jplusphotospectracollection} (see main text). The blue dots and curves represent the photospectra (reflectance points and a cubic spline fit to these points, respectively) computed using the empirical solar colors (G5 set). In red, the results corresponding to the use of theoretical solar colors (2000 ASTM E-490 set). The black curves represent the Bus-DeMeo template for an S-type (left) and a V-type (right).}
\label{fig:spectral_reddening}
\end{figure}

Given these mean variations, we should have in mind that the solar colors choice might induce changes in the slope of some spectra, enough to modify a future taxonomical classification. In order to assess how would this affect taxonomical comparisons, we have selected a subset of photospectra from Appendix \ref{appendix:jplusphotospectracollection} that visually match the templates of an S- (24 objects) and a V-type (4 objects), since these are the easier ones to classify through means of visual inspection.\footnote{The objects that are unambiguosly matched by visual inspection to S-types are 473, 1608, 1677, 1736, 2149, 2498, 2902, 2963, 3259, 3338, 3446, 3863, 4028, 4109, 4703, 5293, 5795, 15419, 18960, 25046, 27082, 46304, 67611, and 102120; V-types are 1709, 2557, 9755, and 18195.} Then, we computed the mean reflectance photospectra for both cases, using the 2000 ASTM E-490 and the G5 solar colors, and compared them to the corresponding taxonomic class templates (S- and V-types). In Figure \ref{fig:spectral_reddening} we can see this systematic difference. Judging by the excesive reddening produced in both cases when using the theoretical colors, compared to the corresponding taxonomic template, we are prone to suggest the empirical colors derived from the G5 set as the optimal choice, as already mentioned.


Nevertheless, we stress that this selection was based in the analysis shown in the present work, and that the perfect choice of the solar colors for the J-PLUS filter system does not exist. Because of this, all the solar colors that we computed in this work are presented in Table \ref{tab:solar_colors}, leaving the final choice to the users of the catalog. In addition, future data releases might provide larger datasets of solar type stars and better statistics, and thorough searches for solar analogs might remove the need of statistical analyses. Thus, we encourage the reader and the catalog users to test other paths and, if possible, improve the solar color computation.


\section{Caveats, issues, and possible solutions}
\label{section4}


\begin{figure}[!ht]
\centering
\includegraphics[width=\hsize]{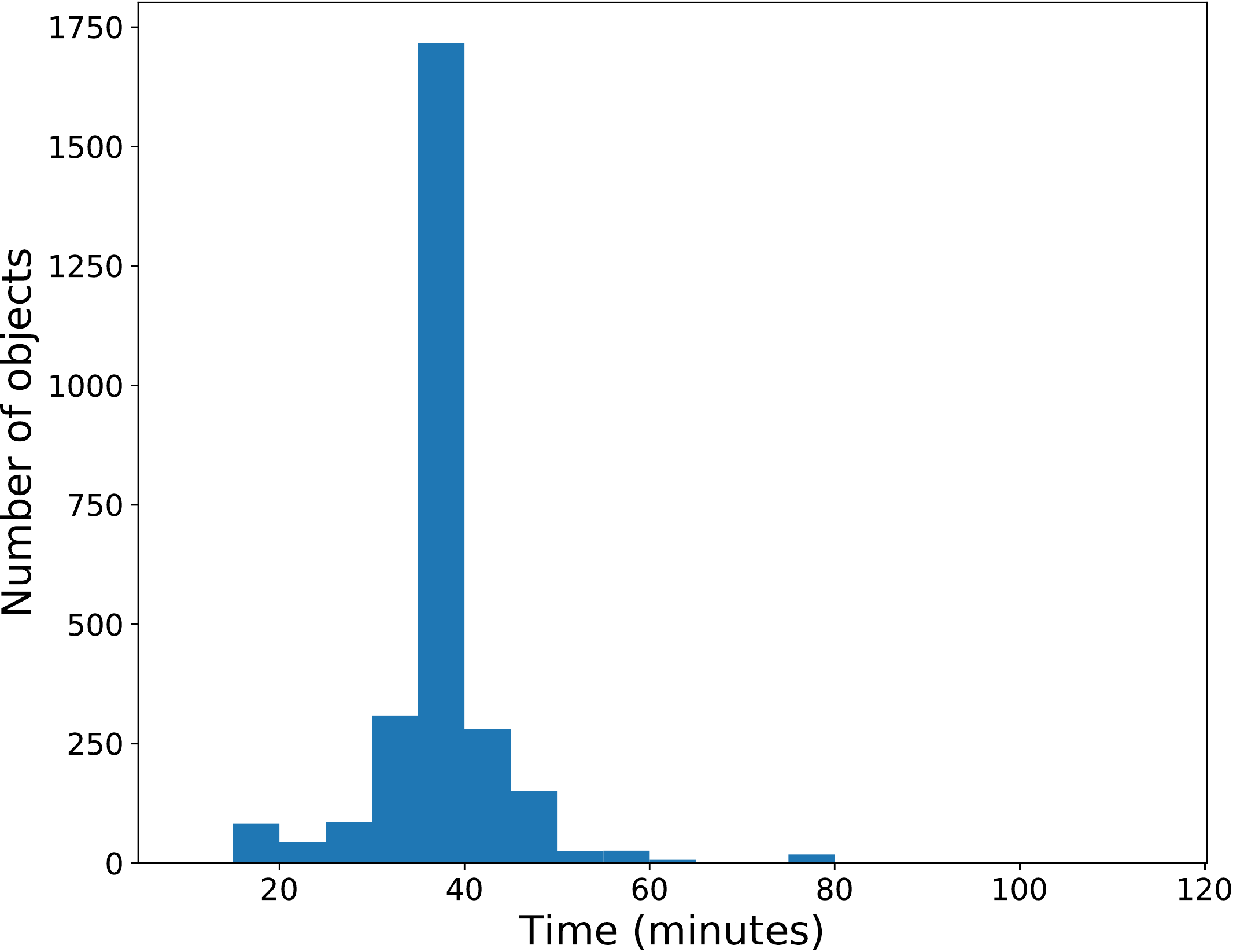}
\caption{Histogram showing the distribution of the exposure times, in minutes, for the asteroids observed in five or more filters. Bin size is 5 minutes. Note that, in some cases, the total exposure time is higher than 60 min. This is due to the choice of the optimal subset of observations for every asteroid (see Section \ref{section2}). However, the majority of the cases are concentrated around the mean value}
\label{fig:mean_exp_times}
\end{figure}


\begin{figure*}[!ht]
\centering
\includegraphics[width=0.815\textwidth]{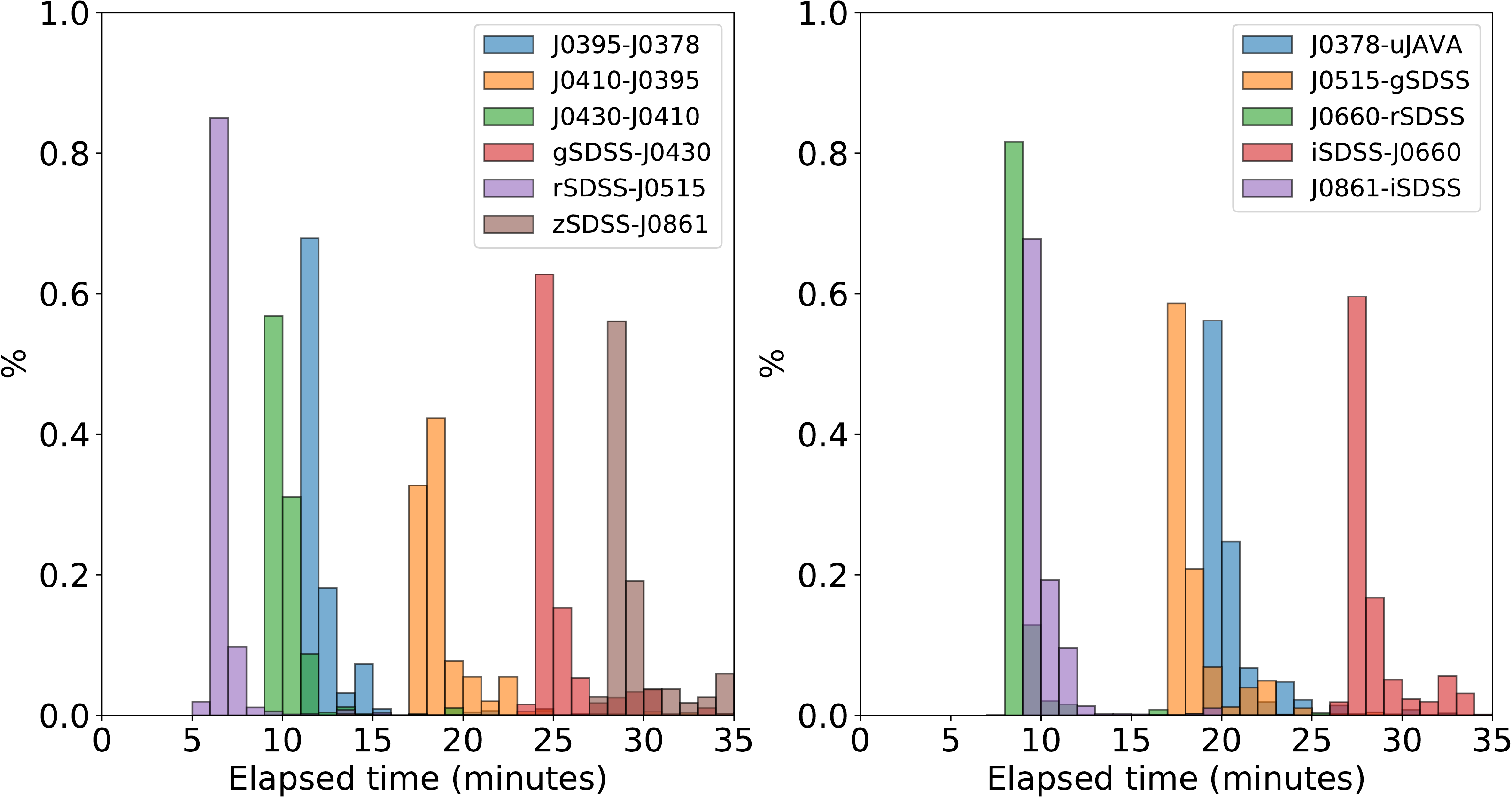}
\caption{Normalized histograms showing the distribution of elapsed times, in minutes, between contiguous filters. Usually, observations in filters at longer wavelengths come (although not immediately) after those at shorter wavelengths, except for the cases of filters $J0410$--$J0395$, $J0430$--$J0410$, $J0515$--$g$, and $i$--$J0660$ (see the filter sequence in Section \ref{section2}). Histograms have been overplotted and separated according to their mean times for a better visual representation. Bin size is 1 minute.}
\label{fig:mean_times_contiguous}
\end{figure*}


\begin{figure}[!ht]
    \centering
    \includegraphics[width=\hsize]{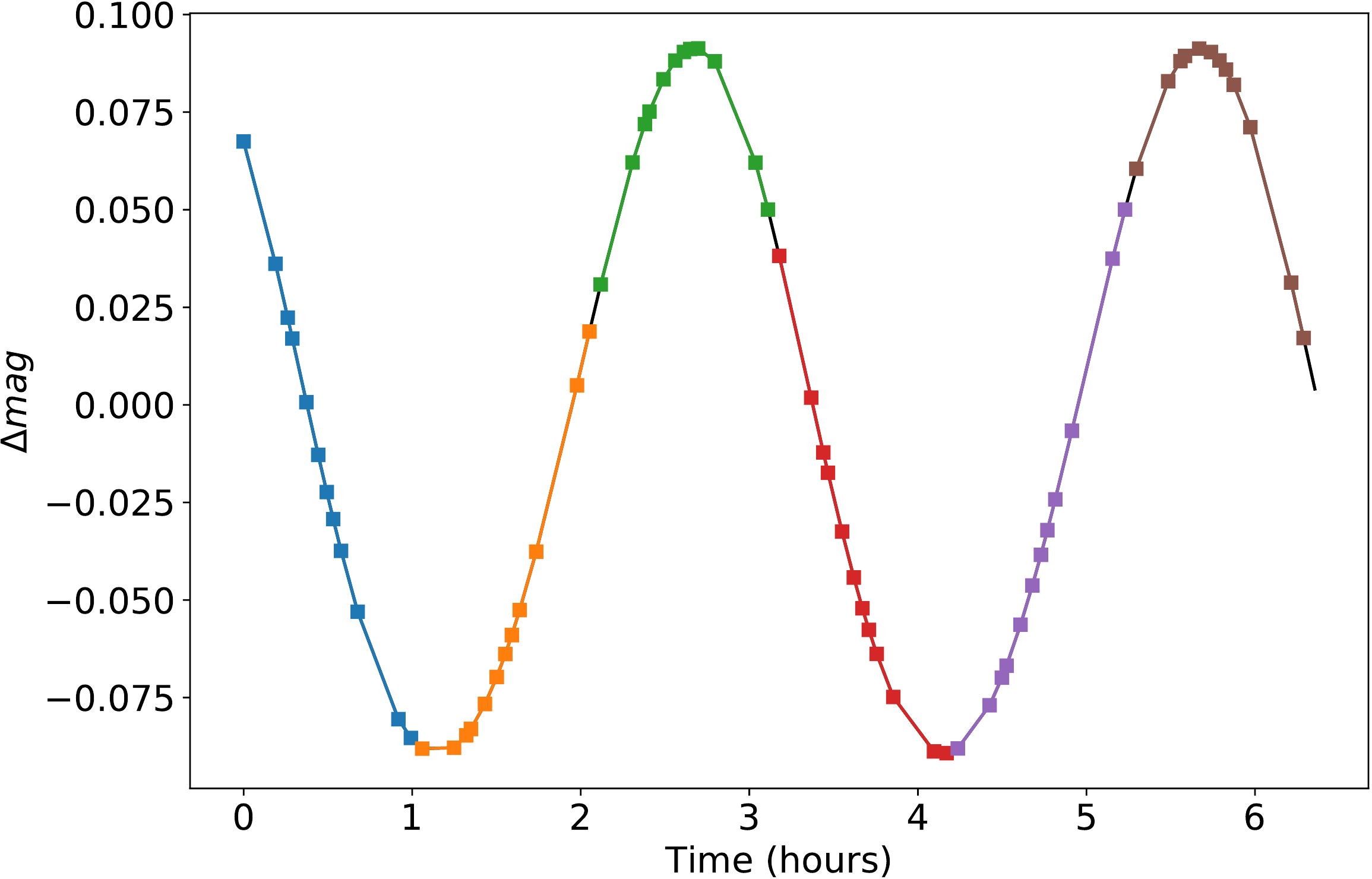}
    \caption{Simulated light curve for an spheroid with axial ratio of 0.7, as observed by J-PLUS. Sets of points with the same color represent the central instant of observations with the 12 J-PLUS filters.}
    \label{fig:rotationvar_lightcurve}
\end{figure}

\begin{figure}[!ht]
    \begin{subfigure}{\columnwidth}
        \includegraphics[width=\hsize]{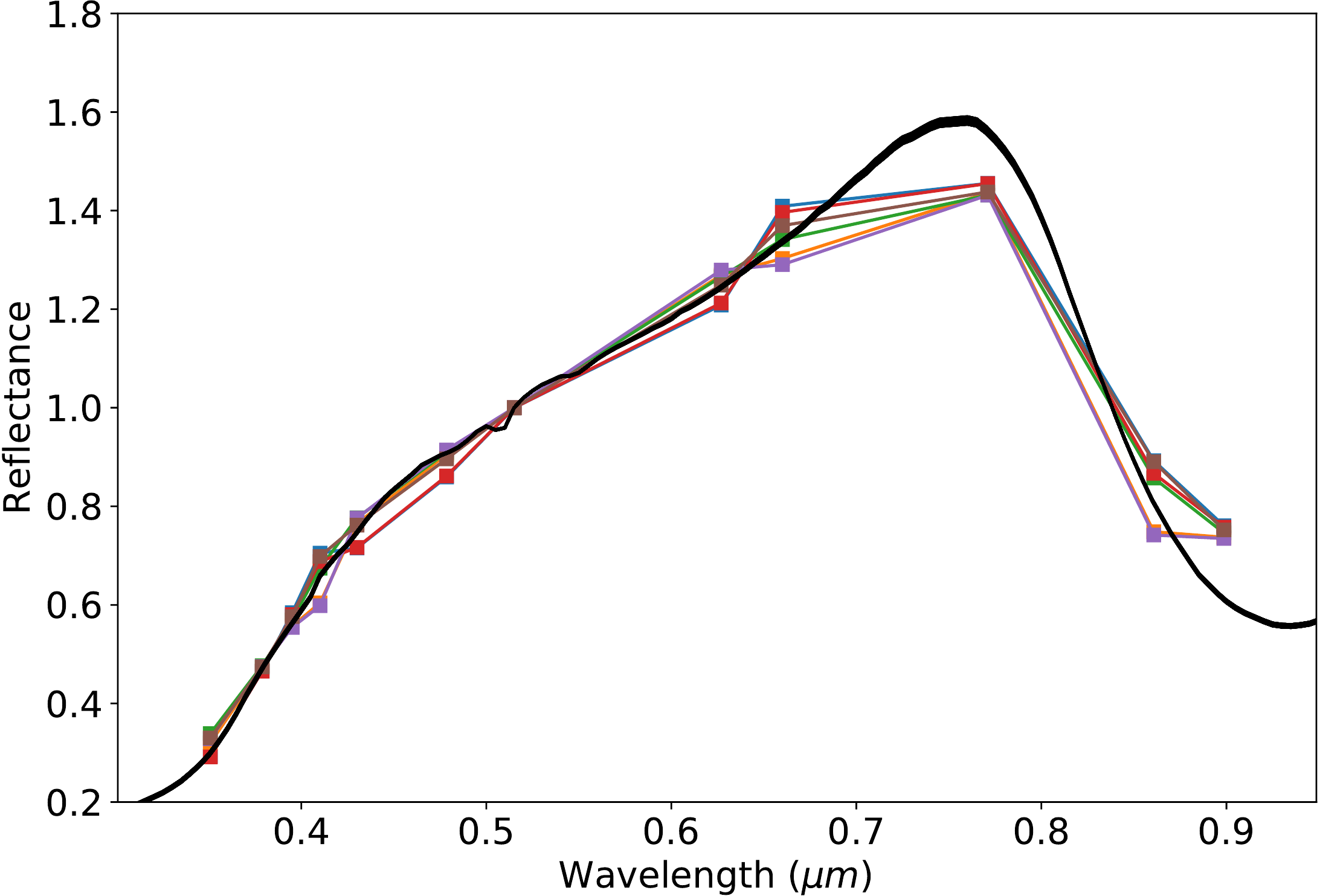}
        \caption{}
    \end{subfigure}
    \begin{subfigure}{\columnwidth}
        \includegraphics[width=\hsize]{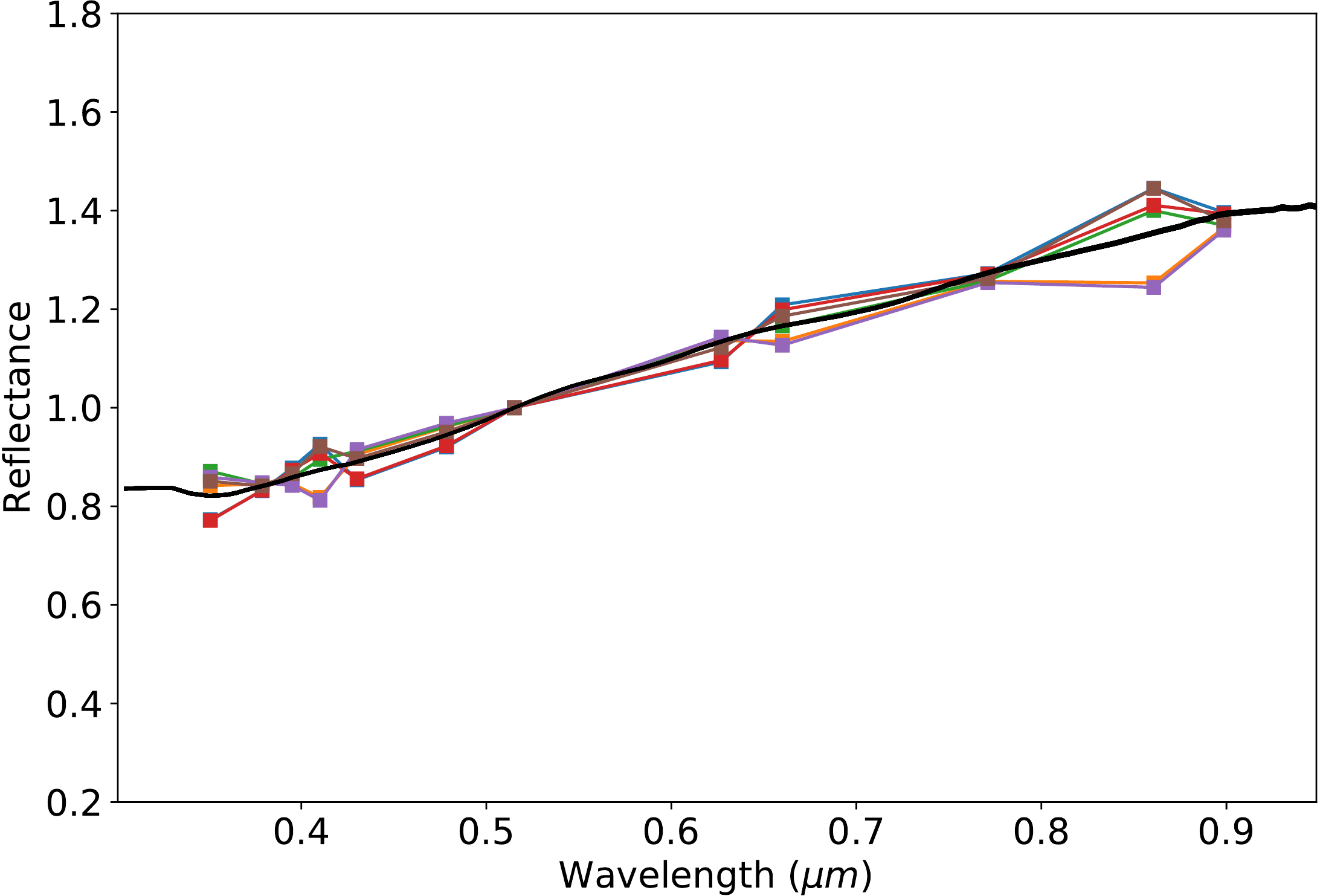}
        \caption{}
    \end{subfigure}
    \caption{(a) Resulting photospectra for the lightcurve in Fig.\ref{fig:rotationvar_lightcurve} using the optical constants of the HED meteorite Allan Hill A76005, with the reflectance spectra of the meteorite shown as a continuous black line. (b) Same as the previous figure, for the optical constants of the Taggish Lake meteorite.}
\label{fig:rotationvar}
\end{figure}

We stress that the spectrophotometry that can be generated using the MOOJa catalog is not exactly equivalent to the spectra that the observed asteroids would present if observed by a spectrograph. Usually, when observing a fixed object in the sky, e.g., a star, obtaining its spectrum or its photospectrum is essentially equivalent, since observations are often carried out on time scales which are significantly shorter than the typical variability time scale of most stars. The same does not hold true for asteroids: minor bodies are normally non-spherical objects, and might present different surface compositions from one region of the asteroid to another. Because of this, and due to their rotation, which is generally in the scale of a few hours, the reflected sunlight varies its intensity along the observation time. If we obtain the spectrum (or photospectrum) of an asteroid using a short enough total exposure time, these fluctuations can be neglected. However, for long exposure times, and depending on the rotation period, shape, and pole orientation of the specific object, the induced reflectance variations might need to be taken into account.

Usually, we refer to exposure times as the total time elapsed since the opening of the shutter of the camera until the moment that the shutter closes, this is, the duration of one exposition. However, because in the case of the MOOJa catalog we are dealing with photometric spectra, we are referring to exposure times as the total time elapsed from the first filter observation to the last, whether the asteroid was observed in the 12 filters or not. In Fig. \ref{fig:mean_exp_times}, we show the total exposure times for all the asteroids observed within the MOOJa catalog. Although some asteroids present more than one hour of total exposure time (given the constraints that we imposed to obtain the spectrophotometry), in the majority of the cases the exposure time is around 40 min, due to the filter observation sequence and to the survey strategy.

The magnitude variation during this interval can be estimated by considering the lightcurves reported by the Asteroid Lightcurve Database\footnote{\url{http://www.minorplanet.info/lightcurvedatabase.html}}. The median rotation period for more than 18\,000 asteroids is $\sim$6.3 h, and the median of the lightcurve amplitudes is $\sim$0.38 magnitudes. Thus, for a lightcurve amplitude of 0.38 magnitudes with a 6.3 h period, and a 40 min interval between the first and last filter measurement, we could estimate a mean uncertainty of $\sim$0.16 magnitudes, which might induce errors around $\sim$15\% when translated to reflectance.

Also, the fact that the filter measurements were not ordered according to the central wavelengths of the filters, as explained in Section \ref{section2}, might introduce non-negligible uncertainties between adjacent points, depending on the time intervals between these contiguous spectrophotometric points (see Fig. \ref{fig:mean_times_contiguous}). In Table \ref{tab:mean_t_contiguous} we show the estimated mean magnitude variations that might be induced by these time separations. Note that the higher estimated variations are those related to the 'jump' from filters $J0861$ to $z$, and from filters $J0660$ to $i$. This is evident by looking at Fig. \ref{fig:jplus_vs_literature}, e.g. for asteroids 462, 1245, 1677, or 2251, where the points corresponding to the $z$ filter are shifted upwards with respect to the $J0861$ points, or the $i$ point is shifted downwards with respect to the $J0660$ point.

It is worth to mention that, in some of the presented plots in Appendixes \ref{appendix:jplus_vs_literature} and \ref{appendix:jplusphotospectracollection}, there seems to persist a number of low-quality observations: asteroid 851, which shows a very red slope; 2730, where there is a point which seems very out of the overall trend; or 6661, 6769, and 9688, which seem to have a very low SNR. In these cases, when the sigma-clipping algorithm was not enough to eliminate these datapoints, some additional procedures that could be used are, for example, filtering data that are highly discrepant with respect to the others (for instance, with respect to adjacent values, preceding and following) or highly discrepant with respect to the overall spectral shape (for instance by fitting the spectra with an appropriate function and identifying outliers with respect to the trend). These options are left unexplored, open for future versions of the catalog.

We have performed some simulations of the theoretical photospectra that would be obtained, as function of asteroid shape, aspect angle, rotational period and superficial composition, in order to assess how rotation could affect the obtained reflectance spectra of asteroids observed by J-PLUS. We consider asteroid shape models as sets of triangular facets. The bidirectional reflectance of each facet is obtained using Hapke's IMSA equation \citep{hapke2005} using a double-lobed Henyey-Greenstein phase function, with the single scatter albedo and phase parameters calculated using geometrical optics from sets of optical constants that vary with wavelength. The integrated reflectance of all facets that are visible and illuminated at a given observational geometry are then calculated for an array of wavelengths, which are then convolved with the band pass of each J-PLUS filter according to the filter sequence and using exposure times that are consistent with the observations. 

Figure \ref{fig:rotationvar_lightcurve} shows the simulated lightcurve for an asteroid with spheroidal shape, an axis ratio of 0.7, a rotation period of 6 hours, and an amplitude of $~0.15\, mag$. Using optical constants for the HED meteorite Allan Hills A76005 \citep{davalos17} and the Taggish Lake meteorite \citep{roush03}, we simulated a V- and a D-type asteroid, respectively. The measured photospectrum would vary depending on the portion of the lightcurve that is observed (see Fig.\ref{fig:rotationvar}).

As expected, the changes induced by rotation in the photospectrum are a function of the amplitude of the portion of the light curve sampled during the observations. The simulated lightcurve has a moderate amplitude ($\sim$0.15); if the amplitude was wider (i.e., the object had a more elongated shape) this variation would be greater, enough to modify spectral points even with small errors (as we see in Figs. \ref{fig:jplus_vs_literature} and \ref{fig:jplus_obs_12filters}). In addition, given that these simulations were performed for two extreme taxonomical classes, such as a V- and a D-type, we expect this effect to induce changes enough to blur the limits between closer classes such as C-, B-, X-, or even S-type asteroids.

The approach that we propose to solve this issue is to take into account the three individual expositions for each filter, using the slopes of the variations in time as some sort of 'partial' lightcurve, in order to correct these rotation-induced magnitude shifts.


\begin{table}
\caption{Mean values and their corresponding errors, in minutes, for the elapsed time between observations corresponding to contiguous filters. The next-to-last column corresponds to the mean magnitude variations estimated using the median values of the lightcurve amplitudes and periods in the Asteroid Lightcurve Database (see main text). The final column represents the percentage of error induced in the reflectance by the variations in magnitude.} 
\label{tab:mean_t_contiguous} 
\centering 
\begin{tabular}{l c c c} 
\hline 
Contiguous filters & $t_{cont}$ (min) & $\Delta$ Mag. & R$_{err}$ (\%) \\
\hline 
$J0378$-$u$ & 20.7 $\pm$ 2.4 & 0.08 & 7\\
$J0395$-$J0378$ & 12.3 $\pm$ 2.1 & 0.05 & 5\\
$J0410$-$J0395$ & 19 $\pm$ 3 & 0.08 & 7\\
$J0430$-$J0410$ & 10.4 $\pm$ 1.3 & 0.04 & 4\\
$g$-$J0430$ & 26 $\pm$ 3 & 0.10 & 9\\
$J0515$-$g$ & 19 $\pm$ 3 & 0.08 & 7\\
$r$-$J0515$ & 6.6 $\pm$ 1.2 & 0.03 & 3\\
$J0660$-$r$ & 8.8 $\pm$ 2.1 & 0.04 & 4\\
$i$-$J0660$ & 29 $\pm$ 4 & 0.12 & 11\\
$J0861$-$i$ & 10.3 $\pm$ 1.4 & 0.04 & 4\\
$z$-$J0861$ & 30 $\pm$ 4 & 0.12 & 11\\
\hline
\end{tabular}
\end{table}

There are also situations in which the reflectance variations between adjacent points are extremely large. These cases probably arise from the fact that the asteroid is passing over, or near, a star in the field at the moment of one of the exposures, or that one of the frames (or all of them) present low data quality. The sigma-clipping algorithm removed some of these outliers. However, even after processing the data, some extreme reflectance variations remain: in order to deal with this issue, the best approach would be to examine the three individual frames obtained for each filter, and discard those in which the asteroid is located too close to a star in the background, or those in which the photometric quality is below some previously imposed threshold. Unfortunately, these individual frames were not available for the first data release, thus, this issue will need to be addressed on a future version of the catalog.


\section{Summary and future work}
\label{section5}

This work focused in presenting to the scientific community the Moving Objects Observed from Javalambre (MOOJa) catalog, a compilation of spectrophotometric data of asteroids observed within the Javalambre Photometric Local Universe Survey in 12 filters covering wavelengths from 0.3 up to 1 $\mu$m. A total of 3\,666 different asteroids were detected. We applied a sigma-clipping algorithm to this initial collection of observations, on the colors that are used to compute spectrophotometry, to clean and remove outliers, and to obtain a more robust dataset. The eventual colors catalog that we provide consists of 3\,122 objects, of which 278 are observed in the 12 filters, i.e., with a complete spectrum. The catalog has been compiled using data only from the first data release of J-PLUS, which covers $\sim$10\% of the sky region intended to be observed. We expect to recover around 18\,000 objects in the full duration of the survey, as well as to discover new ones, obtaining spectrophotometry for these bodies at the same time of the discovery.

We present a new method to select a suitable set of solar colors in the J-PLUS photometric system (out of several options, computed both theoretically and empirically) that will be used in order to compute the photospectra of the asteroids within the MOOJa catalog. This method makes use of those asteroids that have been previously observed by past surveys, and have also been observed by J-PLUS.

We also have introduced the most relevant issues that might prevent us from maximizing the science outcome of the use of the catalog: the most important one, the time interval between observations in contiguous filters; and secondly, flawed observations that cannot be removed with the sigma-clipping algorithm. We propose that the best course of action in order to remove both sources of error is to examine each of the three individual images in every filter: in the first case, correcting the magnitude variation induced by the lightcurve, and in the second, simply removing suboptimal exposures from the computations. Unfortunately, these data were not available as part of the DR1, but are available within the DR2. At the moment there is an ongoing effort to further extend the catalog, which will be updated once all the minor bodies data within the DR2 are recovered and processed.

In addition, we would like to address some interesting points that might serve as a checklist for future works:
\begin{itemize}
\item The first and most immediate analysis should focus on the taxonomic classification of the asteroids observed within the catalog. The fact that the observations are done using 12 filters will provide a huge number of asteroid classifications, and will exponentially increase our knowledge of distribution of different materials throughout the Solar System.
\item Since the number of ultraviolet filters in which asteroids are observed is higher than any previous survey in the field, this might provide new insights into this wavelength region. A novel taxonomic system in the UV should be probed, to enhance our knowledge of minor bodies science within this spectral domain.
\item Given the spectral resolution and the distribution of the filters' central wavelengths, the information on the MOOJa catalog will allow for a more robust definition of the 0.7 $\mu$m absorption band (related to aqueous altered minerals on the surface of the asteroid), than using SDSS data.
\end{itemize}

We would like to mention that there exists another ongoing survey, S-PLUS (Southern Photometric Local Universe Survey, see \citealt{splus19}) that is carried out using an identical duplicate of the telescope used in Javalambre, the T80, located near the summit of Cerro Tololo in central Chile. The aim of this survey is basically the same as J-PLUS, but instead of covering the sky of the northern hemisphere, will cover the southern side. This will provide a similar number of spectrophotometric data of asteroids.

Along these lines, another important survey that might benefit from J-PLUS operations (and viceversa) is the Gaia mission (see \citealt{mignard07, delbo12, tanga12}), since its spectrophotometric system covers basically the same region as J-PLUS. Specifically, the Gaia-BP spectrophotometer covers the UV region, where J-PLUS also offers a total of seven narrow-band and two broad-band filters. Thus, J-PLUS data might be used, for example, to test and calibrate Gaia spectra for minor bodies in the ultraviolet range. The synergies that might arise from combining both datasets are very interesting and should be taken into account.

To summarize, the vast number of minor bodies photospectra that this catalog, and its future versions, will provide, might be the first step into a new era in which large-area surveys will be key to solve unanswered questions in the area of minor bodies of the Solar System, due to the wealth of data (and its quality) that will be available to the community.


\begin{acknowledgements}

{\it DM} acknowledges the CNPq for the support in form of a PCI grant, and also the FAPERJ for the support in the form of a Pós-Doutorado Nota 10 grant.

{\it JMC} acknowledges CPNq for support through a research fellowship.

{\it AAC} acknowledges support from FAPERJ (grant E26/203.186/2016) and CNPq (grants 304971/2016-2 and 401669/2016-5), from the Universidad de Alicante under contract UATALENTO18-02, and from the State Agency for Research of the Spanish MCIU through the
"Center of Excellence Severo Ochoa" award to the Instituto de Astrofísica de Andalucía (SEV-2017-0709)”

{\it MDP} ackowledges funding by the Preeminent Postdoctoral Program (P3) at the University of Central Florida.

{\it JL} acknowledges support from the project AYA2015-67772-R (MINECO).

{\it AG} acknowledges CAPES for the support in the form of a Phd. grant.

{\it MM} acknowledges funding by the European Space Agency under the research contract C4000122918.

{\it ESM and FJE} acknowledges funding by the Spanish State Research Agency (AEI) Projects AYA2017-84089 and MDM-2017-0737 at Centro de Astrobiología (CSIC-INTA), Unidad de Excelencia María de Maeztu.

The present work is based on observations made with the JAST/T80 telescope for the J-PLUS project at the Observatorio Astrofísico de Javalambre in Teruel, a Spanish Infraestructura Cientifico-Técnica Singular (ICTS) owned, managed and operated by the Centro de Estudios de Física del Cosmos de Aragón (CEFCA). Data has been processed and provided by CEFCA's Unit of Processing and Archiving Data (UPAD). Funding for the J-PLUS Project has been provided by the Governments of Spain and Aragón through the Fondo de Inversiones de Teruel; the Aragón Government through the Research Groups E96, E103, and E16\_17R; the Spanish Ministry of Science, Innovation and Universities (MCIU/AEI/FEDER, UE) with grants PGC2018-097585-B-C21 and PGC2018-097585-B-C22; the Spanish Ministry of Economy and Competitiveness (MINECO) under AYA2015-66211-C2-1-P, AYA2015-66211-C2-2, AYA2012-30789, and ICTS-2009-14; and European FEDER funding (FCDD10-4E-867, FCDD13-4E-2685). The Brazilian agencies FAPESP and the National Observatory of Brazil have also contributed to this project.

\end{acknowledgements}

\bibliographystyle{aa} 
\bibliography{biblio.bib} 


\onecolumn
\section*{Appendices}

\appendix

\section{Color vs. color examples}
\label{appendix:colorcolor}
Here we show four color-color distributions of asteroids observed within the MOOJa catalog. In grey, all the observed objects within the colors catalog (that are observed in the corresponding filters). Overplotted, the asteroids (only C-, S-, V-, and X-types) that were previously classified according to their SDSS colors (see \citealt{carvano2010}): C-types, in red; S-types, in black; V-types, in blue; X-types, in green. Top panels are a combination of broad-band-filters-only (left) and narrow-band-filters-only (right). Bottom panels are combinations of both broad and narrow-band filters. It is easy to separate between, at least, C- and S-types using the combinations in the top- and bottom-left panels.
\FloatBarrier
\begin{figure*}[h]
\centering
\includegraphics[width=\hsize]{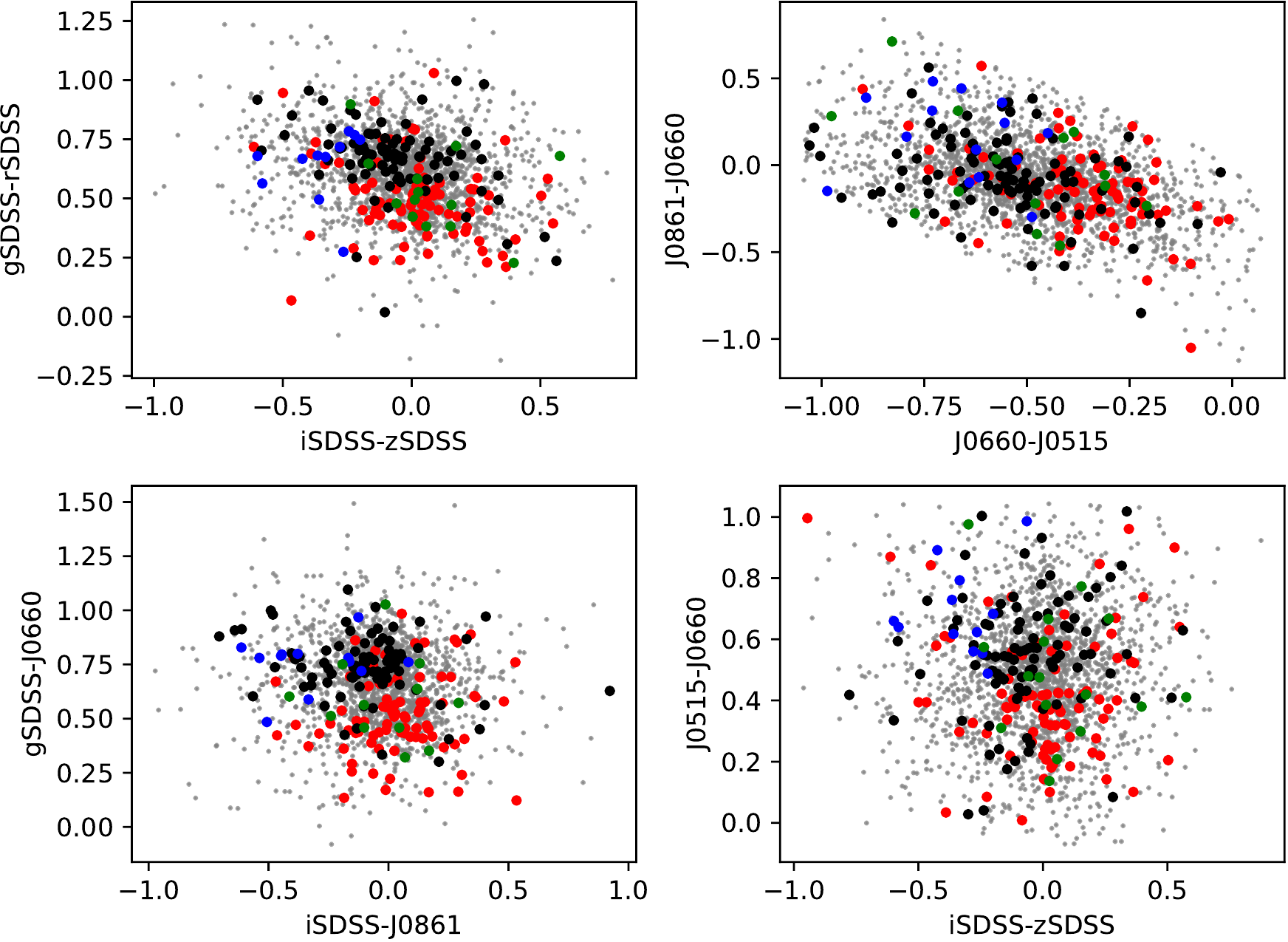}
\caption{Color-color distributions for some filter combinations.}
\label{fig:color_color_plots}
\end{figure*}
\FloatBarrier

\clearpage

\section{J-PLUS spectra vs. literature spectra}
\label{appendix:jplus_vs_literature}
Photospectra in the 0.3-1.0 $\mu$m region of the 33 asteroids within the MOOJa catalog which are also available in the literature. The J-PLUS spectrophotometry (computed using the G5 solar colors, see Sect.\ref{section3}) is represented in blue, and the literature spectra, overplotted, are represented as follows: ECAS, orange right-pointing triangles, $\vartriangleright$; SMASSI, green circles, $\circ$; SMASSII, black squares, $\Box$; S3OS2, magenta left-pointing triangles $\vartriangleleft$; 24CAS, gray down-pointing arrows $v$; Sawyer, cyan diamonds, $\Diamond$; PRIMASS, red crosses, $\times$. All photospectra are normalized at 0.515 $\mu$m, which is the central wavelength of the $J0515$ filter, which, among the central wavelengths of all the filters, is the nearest one to 0.55 $\mu$m, the most widely used value to normalize visible asteroid's spectra.
\FloatBarrier

\begin{figure*}[h]
\centering
\includegraphics[width=0.815\textwidth]{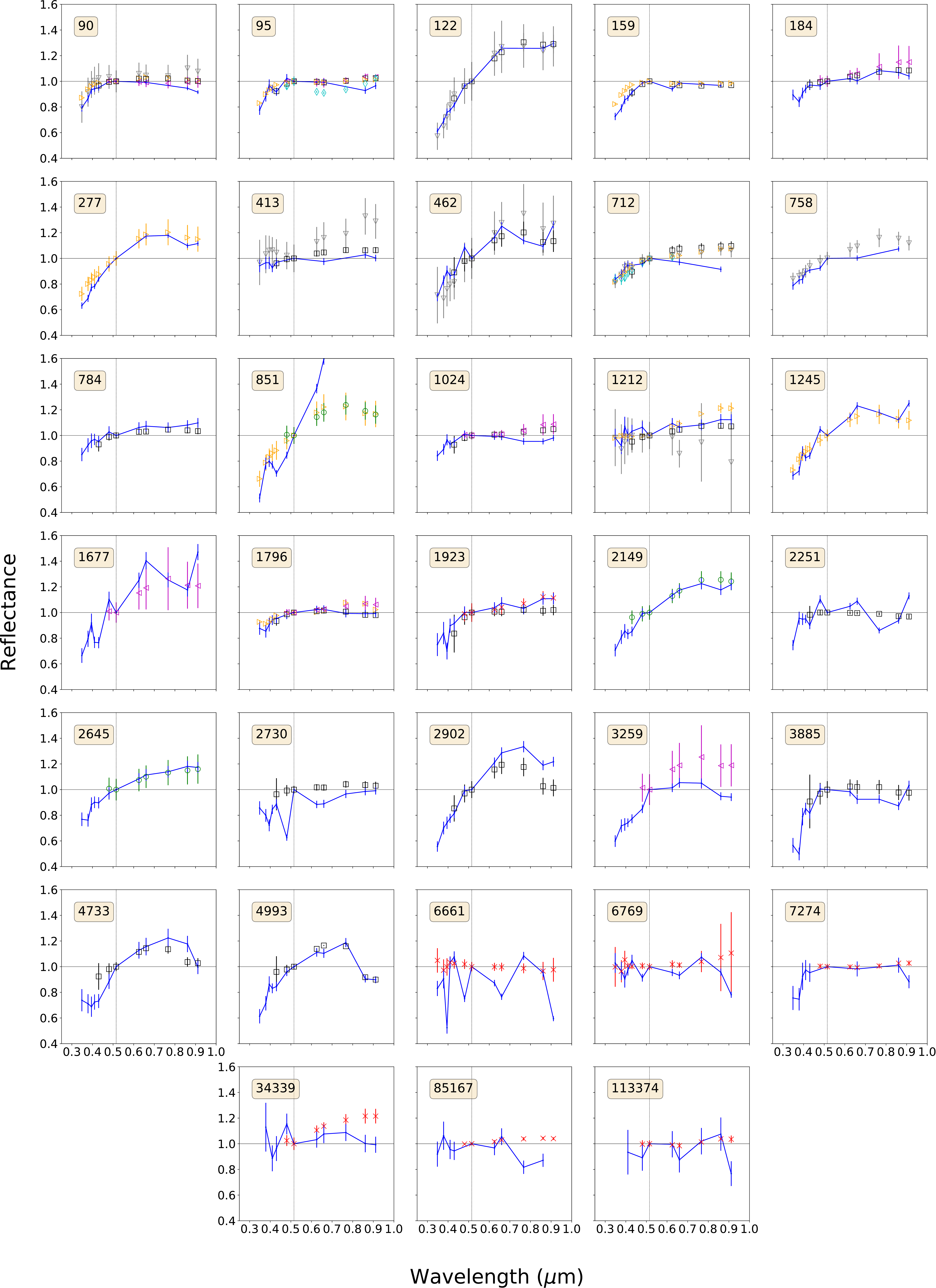}
\caption{Comparison between photospectra of J-PLUS asteroids and visible literature spectra.}
\label{fig:jplus_vs_literature}
\end{figure*}
\FloatBarrier

\clearpage

\section{J-PLUS photospectra collection}
\label{appendix:jplusphotospectracollection}
Photospectra in the 0.3-1.0 $\mu$m region of 74 asteroids within the MOOJa catalog. These objects have been detected by J-PLUS in all the 12 filters and with $(m_f - m_{J0515})_{err}<0.1$ magnitude units. The dots represent the J-PLUS data, and the solid curve is a cubic polynomial fit, for a better visualization of the overall spectral shape.
\FloatBarrier

\begin{figure*}[h]
\centering
\includegraphics[width=0.815\textwidth]{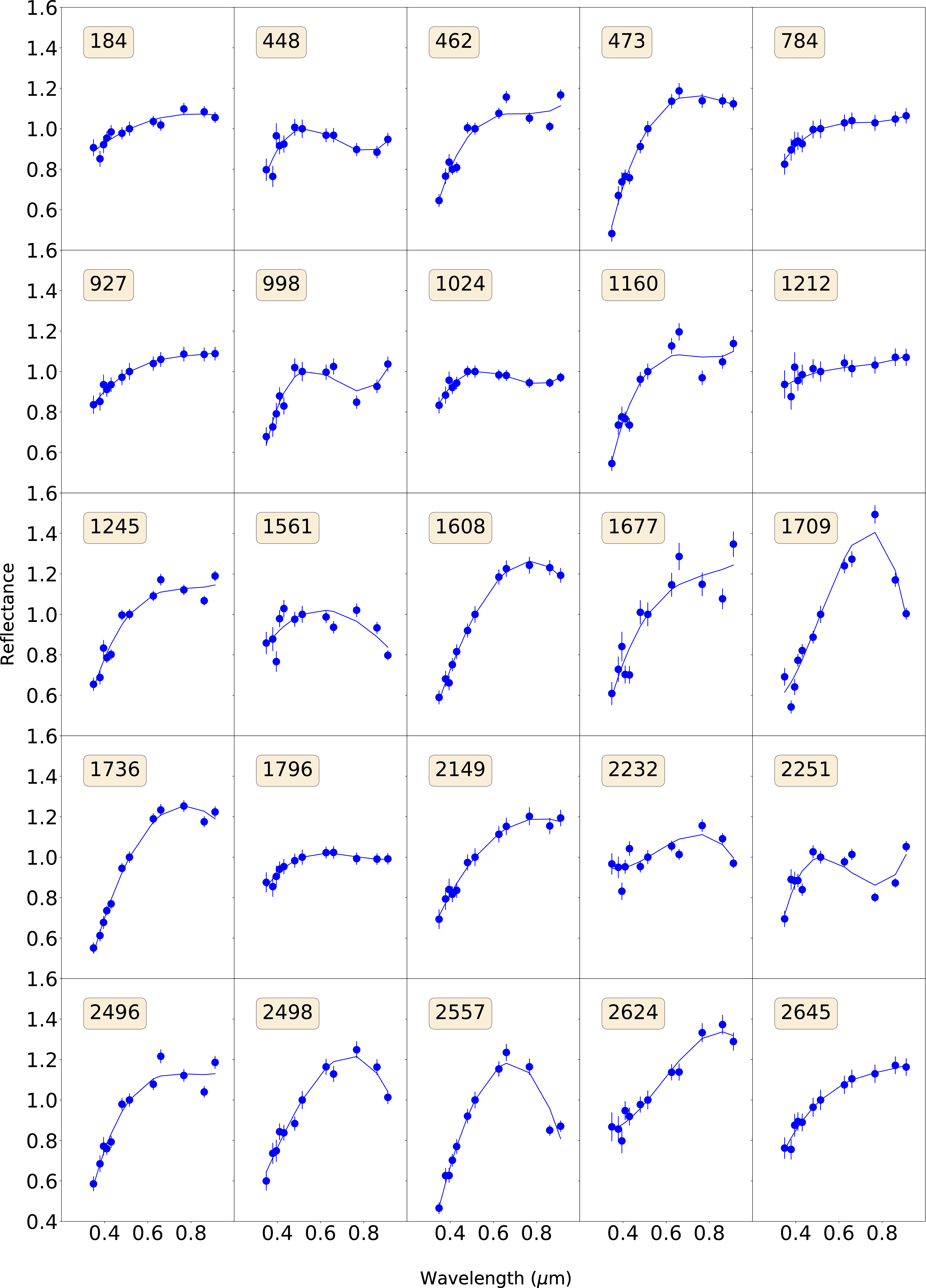}
\caption{Photospectra of asteroids detected in the MOOJa catalog.}
\label{fig:jplus_obs_12filters}
\end{figure*}
\FloatBarrier

\begin{figure*}[h]
\ContinuedFloat
\centering
\includegraphics[width=0.815\textwidth]{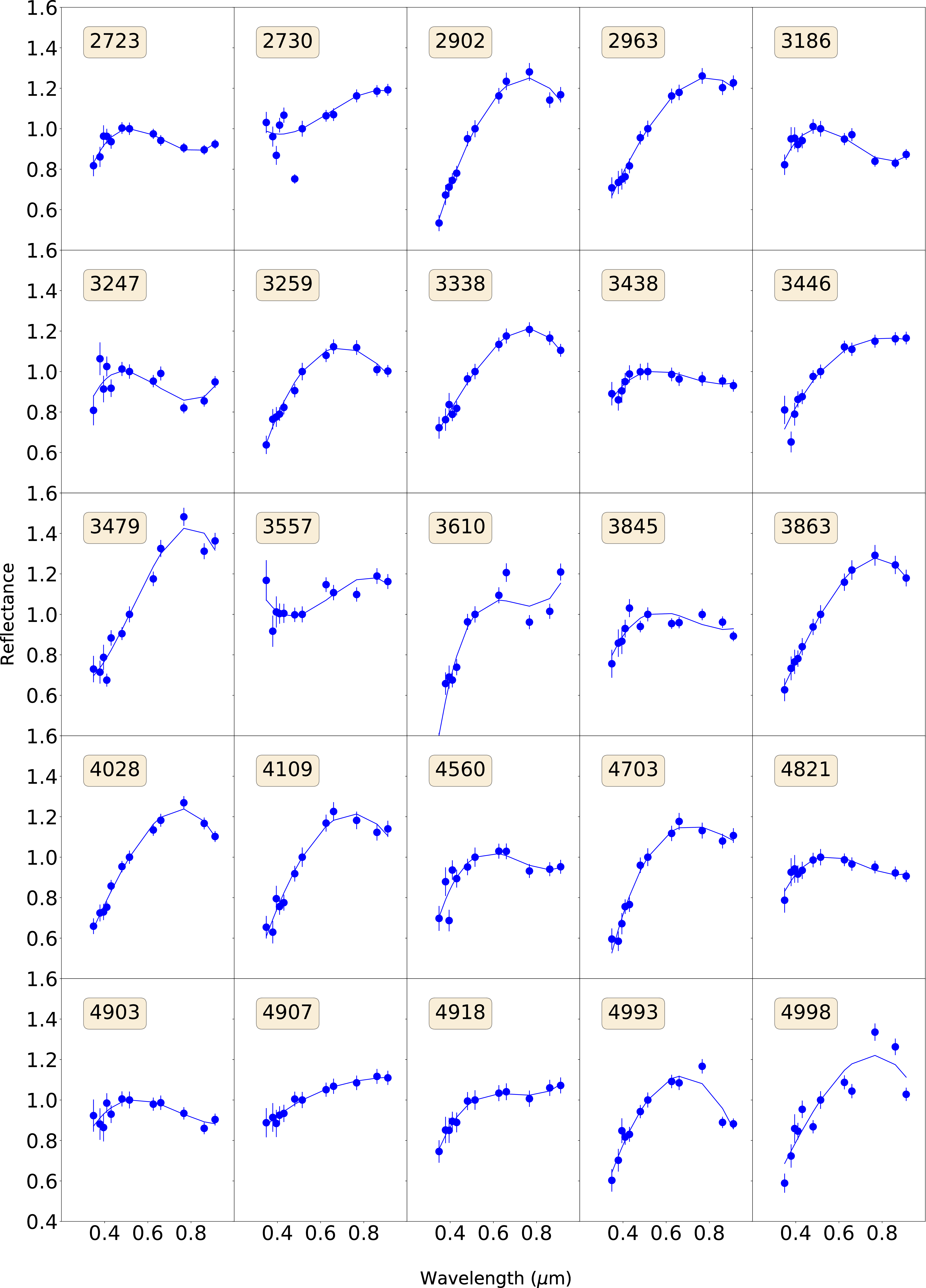}
\caption[]{Photospectra of asteroids detected in the MOOJa catalog (continued).}
\end{figure*}
\FloatBarrier

\begin{figure*}[h]
\ContinuedFloat
\centering
\includegraphics[width=0.815\textwidth]{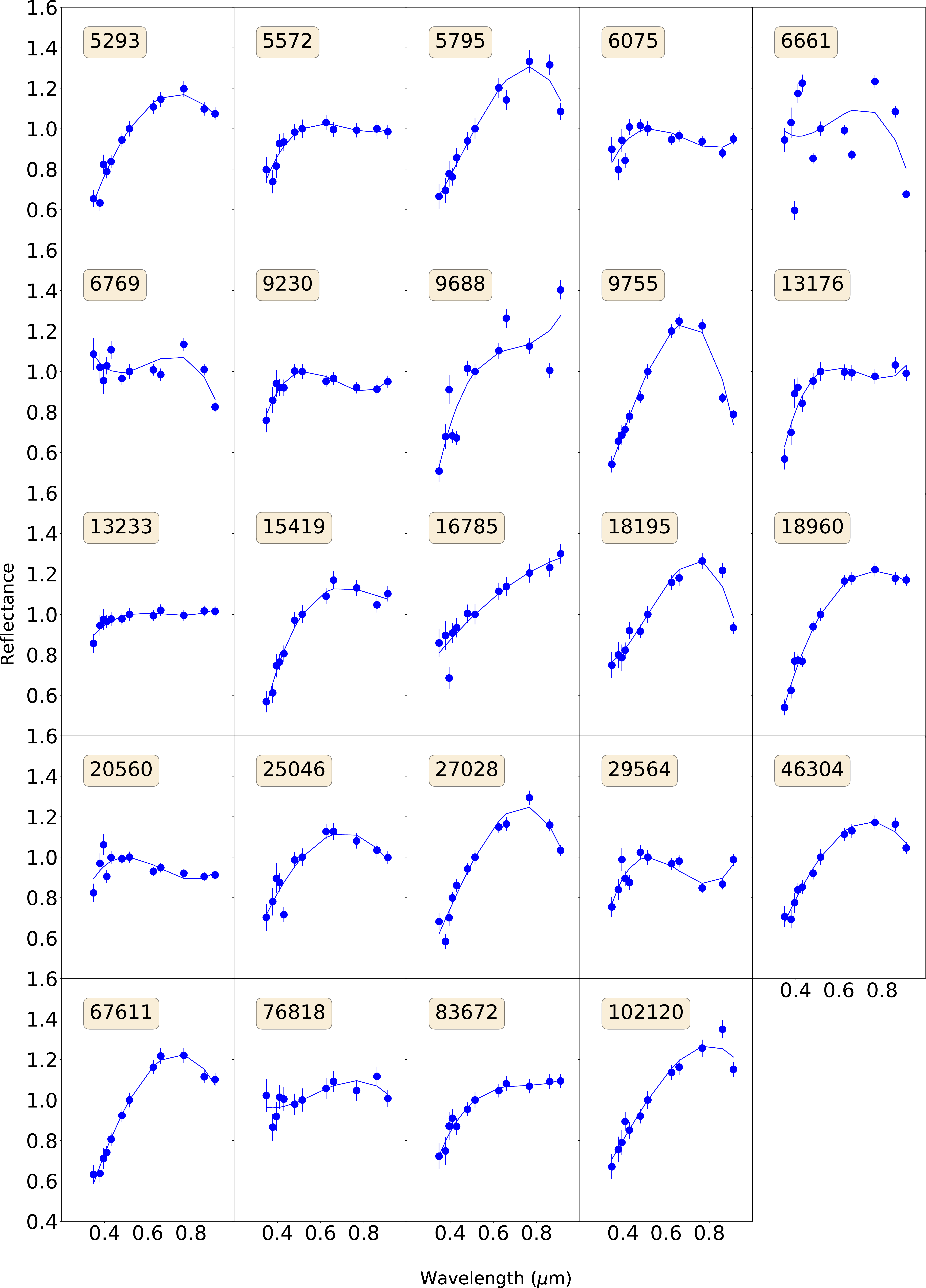}
\caption[]{Photospectra of asteroids detected in the MOOJa catalog (continued).}
\end{figure*}
\FloatBarrier

\clearpage

\end{document}